\newcommand\oiii{\hbox{[O{$\,$\scriptsize$\rm III$}]}}
\newcommand\siii{\hbox{[Si{$\,$\scriptsize$\rm II$}]}}
\newcommand\oi{\hbox{[O{$\,$\scriptsize$\rm I$}]}}

\newcommand\cii{\hbox{C{$\,$\scriptsize II}}}
\newcommand\civ{\hbox{C{$\,$\scriptsize IV}}}
\newcommand\heii{\hbox{He{$\,$\scriptsize II}}}
\newcommand\nv{\hbox{N{$\,$\scriptsize V}}}
\newcommand\niv{\hbox{N{$\,$\scriptsize IV}]}}
\newcommand\siiii{\hbox{Si{$\,$\scriptsize III}]}}
\newcommand\siiv{\hbox{Si{$\,$\scriptsize IV}}}
\newcommand\oiv{\hbox{O{$\,$\scriptsize IV}]}}
\newcommand\ciii{\hbox{C{$\,$\scriptsize III}]}}
\newcommand\feii{\hbox{Fe{$\,$\scriptsize II}}}
\newcommand\feiii{\hbox{Fe{$\,$\scriptsize III}}}
\newcommand\mgii{\hbox{Mg{$\,$\scriptsize II}}}
\newcommand\alii{\hbox{Al{$\,$\scriptsize II}}}
\newcommand\aliii{\hbox{Al{$\,$\scriptsize III}}}

\newcommand\lya{Ly$\alpha$}

\newcommand\kms{\ifmmode{\rm km}\,{\rm s}^{-1}\else km$\,$s$^{-1}$\fi}
\newcommand\ergscm{$10^{-17}$ erg s$^{-1}$ cm$^{-2}$}
\newcommand\ergscmA{$10^{-17}$ erg s$^{-1}$ cm$^{-2}$ \AA$^{-1}$}

\documentclass[useAMS,usenatbib,onecolumn]{mn2e}
\usepackage{comment,amssymb,graphicx,color,longtable,lscape}
\title[Candidate Type II Quasars]{Candidate Type II Quasars at $2 < z
< 4.3$ in the Sloan Digital Sky Survey III} 

\author[R. Alexandroff et al.]{Rachael
Alexandroff$^{1,2}$\thanks{E-mail: rmalexan@pha.jhu.edu}, Michael
A. Strauss$^{2}$, Jenny E. Greene$^{2}$, Nadia L. 
  Zakamska$^{1}$, 
\newauthor
Nicholas P. Ross$^{3}$, W. N. Brandt$^{4,5}$, Guilin Liu$^{1}$, Paul
S. Smith$^6$, 
Jian Ge$^7$,
Fred Hamann$^7$, 
\newauthor
Adam D. Myers$^8$, 
Patrick Petitjean$^9$,
Donald P. Schneider$^{4,5}$,
Hassen Yesuf$^{10}$,
\newauthor
Donald G. York$^{11}$\\
$^{1}$Center for Astrophysical Sciences, Department of Physics and Astronomy, Johns Hopkins University, Baltimore, MD 21218, USA.\\
$^{2}$Department of Astrophysical Sciences, Princeton University, Princeton, NJ 08544, USA.\\
$^{3}$Lawrence Berkeley National Laboratory, 1 Cyclotron Road, Berkeley, CA 92420, USA.\\
$^{4}$Department of Astronomy and Astrophysics, The Pennsylvania State
  University, 525 Davey Laboratory, University Park, PA 16802, USA.\\ 
$^{5}$Institute for Gravitation and the Cosmos, The Pennsylvania State
  University, 
University Park, PA 16802, USA.\\
$^{6}$Steward Observatory, University of Arizona, 933 North Cherry
Avenue, Tucson, AZ 85721, USA.\\
$^7$Department of Astronomy, University of Florida, Gainesville, FL 32611-2055, USA.\\
$^8$Department of Physics and Astronomy, University of Wyoming, Laramie, WY
82072, USA.\\
$^9$UPMC-CNRS, UMR7095, 
Institut d'Astrophysique de Paris, 
98bis Boulevard Arago, F-75014, Paris, France.\\
$^{10}$UCO/Lick Observatory, University of California, Santa Cruz, 1156 High Street,
Santa Cruz, CA 95064, USA.\\
$^{11}$Department of Astronomy and Astrophysics and the Enrico Fermi Institute, University of Chicago, 
5640 South Ellis Avenue, Chicago, IL\\ 
60637, USA.\\
}

\begin{document}

\date{}

\pagerange{\pageref{firstpage}--\pageref{lastpage}} \pubyear{2013}

\maketitle

\label{firstpage}

\begin{abstract}
At low redshifts, dust-obscured quasars often have strong yet narrow
permitted lines in the rest-frame optical and ultraviolet, excited by
the central active nucleus, earning the designation Type II quasars.  
We present a sample of 145 candidate
Type II quasars at redshifts between 2 and 4.3, encompassing the
epoch at which quasar activity peaked in the universe.  These objects,
selected from the quasar sample of the Baryon Oscillation
Spectroscopic Survey of the Sloan Digital Sky Survey III, are
characterized by weak continuum in the rest-frame ultraviolet (typical
continuum magnitude of $i\approx 22$) and strong lines of \civ\ and
\lya, with Full Width at Half Maximum less than 2000 \kms.  The
continuum magnitudes correspond to an absolute magnitude of $-23$ or
brighter at redshift 3, too bright to be due exclusively to the host
galaxies of these objects.  Roughly one third of the objects are
detected in the shorter-wavelength bands of the WISE survey; the
spectral energy distributions (SEDs) of these objects appear to be
intermediate between classic Type I and Type II quasars seen at lower
redshift.  Five objects are detected at rest frame 6\micron\ by
Spitzer, implying bolometric luminosities of several times $10^{46}$ erg
s$^{-1}$.  We have obtained polarization measurements for two objects;
they are roughly 3\% polarized.  We suggest that these objects are
luminous quasars, with modest dust extinction ($A_V \sim 0.5$ mag),
whose ultraviolet continuum also includes a substantial scattering
contribution.  Alternatively, the line of sight to the central engines
of these objects may be obscured by optically thick
material whose covering fraction is less than unity.  
\end{abstract}


\section{Introduction}

In standard unification models, many of the observed properties of
Active Galactic Nuclei (AGN) can be explained by differences in
viewing angle \citep{1993ARAA..31..473A, 1995PASP..107..803U}.  In
these models, the accretion disk of the supermassive black hole (SMBH)
is surrounded by a torus of gas and dust, which, when oriented along
the line of sight, obscures emission from the region around the SMBH
at optical, ultraviolet and soft X-ray wavelengths. Because this gas and dust
does not cover all $4\,\pi$ steradians around the central engine, gas
in the host galaxy above and below the torus is illuminated by the engine,
giving rise to strong narrow high-ionization emission lines
(Full Width at Half Maximum (FWHM) $< 2000$ \kms) and weak continua
\citep{2003AJ....126.2125Z}
in the rest-frame optical spectra.  Such
objects are classified as Type II based on their optical spectra, in
contrast to Type I objects which show strong ultraviolet continua and
broad permitted lines \citep{1971Ap......7..231K,1974ApJ...192..581K}.  Type II AGN tend to show a high ratio of IR to
optical light, to have hard X-ray spectra, and to be strongly polarized,
consistent with the obscuring torus hypothesis \citep{antonucci85,2002ApJ...571..218N,
2002ApJ...569...23S,2003AJ....126.2125Z,2005ARA&A..43..827B}.
However, \citet{1988ApJ...325...74S}, \citet{2001ApJ...555..719C} and \citet{2006ApJS..163....1H} argue
that Type I and Type II quasars represent different phases in quasar
evolution: in their models, all
quasars pass through an obscured phase before outflows from the AGN
and central star formation expel the obscuring material.  

The comoving space density of luminous Type I quasars peaked at
redshifts 2-3 \citep{1995AJ....110...68S,2006AJ....131.2766R,Ross12b}, although the
demographics of luminous obscured quasars at this epoch are
poorly understood.  While it is straightforward to identify 
unobscured quasars as ultraviolet-excess sources in multi-band optical
surveys
\citep{1965ApJ...141.1560S,1986ApJS...61..305G,2006AJ....131.2766R},
a complete census of AGN is challenging at visible 
wavelengths  
alone, given that an appreciable fraction of the quasar population is
obscured by dust. 
Astronomers have used searches in a variety of wavebands to identify
obscured quasars \citep[see, for example,][]{2005ARA&A..43..827B,
  2005ApJ...631..163S,   2005AJ....129.1198H,
2007A&A...463...79G, 2009ApJ...696..110T, 
2010MNRAS.402.1081V, 2011ApJ...736...56B, 2012ApJ...758..129X, 2012ApJ...748..142D, 
2013ApJ...772...26A, 2012ApJ...753...30S, 2013A&A...556A..29M}.      
  
The integral of the quasar luminosity function, with appropriate
efficiency factors, approximately matches the present-day mass
function of SMBH \citep{Soltan82,YuTremaine02,Marconi04} indicating
that black holes accrete much of their mass during a luminous phase as
quasars. This argument has profound implications for understanding the
growth of black holes and their role in galaxy evolution, but
improving this calculation requires good measurements of quasar
demographics, including the obscured quasar fraction, as a function of
redshift and luminosity. This goal remains elusive: surveys at
different wavelengths often find discrepant results
\citep{2010ApJ...714..561L}. At $z<0.8$,  \citet{2008AJ....136.2373R}
find that the ratio of optically-selected Type II to Type I luminous
quasars is at least 1:1, while X-ray studies
\citep[for example, ][]{2003ApJ...598..886U,2008A&A...490..905H} place the
value at $\sim$ 3:1 for low-luminosity active nuclei, but $<1:1$ for
high-luminosity quasars. Using X-ray data, 
\citet{2000Natur.404..459M} and \citet{2006ApJ...645...95H} suggest that the
obscured fraction remains constant or even increases with redshift
\citep[see reviews by][for further discussion]{2011ApJ...736...56B,
2012AinA}.  
    
High-redshift radio-loud Type II quasars have been studied for decades 
\citep[see, for example, the review paper by][]{1993ARAA..31..639M} but their radio-quiet
counterparts in the optical, IR and X-ray have been harder to find.
Deep Mid-IR
\citep[e.g.][]{2005Natur.434..738A,2005ApJ...631..163S,2008ApJ...677..943D,
2010ApJ...713..503C,2012ApJ...748..142D,2012ApJ...753...30S} and X-ray
\citep[e.g.][]{2009ApJ...693.1713T,
2011A&A...526L...9C,2012ApJ...752...46L, 2013ApJ...763..111V} surveys
tend to cover small solid angles, and are thus not sensitive to rare
luminous objects.  Moreover, indications of obscuration do not always
agree between different wavebands
\citep[e.g.,][]{2005AJ....129..578B,2012ApJS..201...30C,2012arXiv1205.0033J}.
For example, about 50\% of X-ray identified Compton-thick objects show
broad emission lines in their optical spectra
\citep[e.g.,][]{2009MNRAS.399.1553V}.
  
The Sloan Digital Sky Survey (SDSS;
\citealt{York00,2011AJ....142...72E}) has covered almost 1/3 of the Celestial Sphere in both
visible-light imaging and spectroscopy to a depth at which significant
numbers of high-redshift quasars are found
\citep{Richards02}. \citet{2003AJ....126.2125Z} and
\citet{2008AJ....136.2373R} 
selected high-luminosity Type II objects with $z < 0.83$ among SDSS
spectra of galaxies \citep{Strauss02} and Faint Images of the Radio
Sky at Twenty-cm (FIRST) radio sources
\citep{1995ApJ...450..559B}.  These objects were identified by their
strong narrow emission lines (FWHM less than 1000 \kms\ in most cases)
and weak continuum; \oiii5008\AA\ was used as a crude proxy for
bolometric luminosity \citep{Heckman05}.  Additional observations of
these objects, including spectropolarimetry
\citep{2005AJ....129.1212Z,2006AJ....132.1496Z} and mid-infrared
photometry and spectroscopy \citep{Zakamska08} demonstrated that these
objects were indeed highly luminous obscured quasars, with a space
density (at least to $z \sim 0.8$) comparable to unobscured quasars
\citep{2008AJ....136.2373R}.

Searches for counterparts at higher redshift were not successful in
the SDSS-I/II data; narrow-line objects typically had strong continua and
extensive Fe emission, showing them to be high-redshift analogs of
Narrow Line Seyfert I galaxies (NLS1; \citealt{1985ApJ...297..166O,Williams01}).  However,
the Baryon Oscillation 
Spectroscopic Survey (BOSS; \citealt{Dawson13}) of SDSS-III \citep{2011AJ....142...72E}
targets quasars two
magnitudes fainter than SDSS-I/II did \citep{2012ApJS..199....3R}, probing to
continuum levels at which Type II quasar candidates at high redshift begin to
appear.

  In this paper, we describe a class of high-redshift ($z > 2$) Type II
  quasar candidates identified by their characteristic spectra from
  BOSS.  We have found 452 candidates with redshifts in the range $2.03 < z < 4.23$
in the SDSS-III 
Data Release 9 \citep[][hereafter referred to as
  DR9]{DR9}. We describe the relevant SDSS data in \S~\ref{sec:SDSS}
and the selection of our candidates in \S~\ref{sec:selection}.  The
properties of these objects in SDSS data are described in
\S~\ref{sec:optical}, and we match against other datasets in
\S~\ref{sec:other_bands}. We discuss our results in \S~\ref{sec:discussion},
and conclude in \S~\ref{sec:conclusions}. 
We assume a flat $\Lambda$CDM cosmology with $\Omega_m$ = 0.26,
$\Omega_{\Lambda}$ = 0.74 and $h$ = 0.71 \citep{2007ApJS..170..377S}
throughout this paper. We use AB magnitudes consistently in this
paper. 

\section{SDSS Observations and Data Processing}
\label{sec:SDSS}

The SDSS has been in routine operation
since 2000.  It uses the dedicated 2.5-meter wide-field Sloan
Foundation Telescope at Apache Point Observatory in New Mexico
\citep{2006AJ....131.2332G}, carrying out both imaging \citep{Gunn98}
and spectroscopy.  The two optical spectrographs were
upgraded in 2009 for the BOSS survey \citep{2013AJ....146...32S}; each is fed by 500
optical fibers  with 2$''$ optical diameter, yielding
spectrophotometrically calibrated spectra from 3600\AA\ to 10,400\AA
\footnote{As described in \citet{Paris12} and \citet{Dawson13}, fibers
for quasar candidates in BOSS
were often offset from their fiducial positions to maximize the throughput in the
blue given differential chromatic refraction.  However, the
spectrophotometric standard stars were observed without these offsets,
causing systematic errors in the spectrophotometric
calibration of quasars of up to 40\%.}, with resolution
$\lambda/\Delta \lambda \approx 1800$.
BOSS is designed to measure the baryon oscillation feature in the
clustering of galaxies \citep{Anderson12} and the \lya\ absorption spectra
of quasars \citep{2013A&A...552A..96B}. 

BOSS data were first made public in the SDSS DR9, containing spectra of 536,000 galaxies
and 102,000 quasars over 3275 deg$^2$.  Because of the need to observe
the Ly$\alpha$ 
forest, quasars are targeted in the region of color space where
objects with $2.15 < z < 3.5$ are expected to lie \citep{2012ApJS..199....3R}.
This is a challenging task, because the broad-band colors of $z \sim 2.7$
quasars are similar to those of much more numerous F and A stars
\citep{Fan99}.  The quasar candidates are selected to a Point Source
Function magnitude
limit of $g \leqslant 22.0$ or $r \leqslant 21.85$ (after correction
for \citealt{SFD} extinction).  All spectra are processed with a common pipeline \citep{Bolton12}; 
\citet{Paris12} report that almost 97\% of quasar targets have spectra
of sufficient signal-to-noise ratio (S/N) to measure a reliable
redshift.  

At redshifts above $z \sim 1$, the H$\alpha$, H$\beta$ and \oiii\
emission lines commonly used as diagnostics for identifying Type II
quasars at optical wavelengths no longer fall within the wavelength
coverage of the BOSS spectrograph.  Instead, we used the widths of the
Ly$\alpha$ (1216\AA) and \civ\ (1549\AA) emission lines as diagnostics
of candidate Type II objects, cutting at FWHM $<$ 2000~\kms\ (see,
e.g., \citealt{2003AJ....126.2125Z, 2005AJ....129.1783H,
2007ApJ...666..757S}).  Given the BOSS spectral coverage, we can
measure Ly$\alpha$ cleanly at redshifts $z \gtrsim 2.0$.  Line widths
are measured directly by the BOSS pipeline, from single Gaussian fits
to the Ly$\alpha$ and \civ\ lines.

\section{Sample Selection Criteria}
\label{sec:selection}

Heavily dust-obscured quasars are expected to have strong, narrow emission lines
atop a relatively weak continuum, which is a combination of the light
from the host galaxy and the light from the hidden quasar scattered by
the interstellar material in the host. The classical definition of
optically selected Type II active nuclei largely focuses on the widths
and ratios of emission lines, which is the approach that
\citet{2008AJ....136.2373R} used in selecting $\sim 900$ Type II
quasars at $z<0.8$ from SDSS-I/II data.  But because obscured quasars
are faint at rest-frame optical and ultraviolet wavelengths, previous
searches for Type II quasars at higher redshifts using the SDSS-I/II
data have been largely unsuccessful.  The redshift range between 0.8
and 2.0 is particularly difficult because none of the strong emission
lines characteristic of Type II quasars appear at the optical
wavelengths.  Only one candidate, SDSS~J085600.88+371345.5 at $z=1.02$
was identified in the original SDSS quasar sample as satisfying all
our emission line criteria. This object was also identified as an
obscured quasar candidate by \citet{gilli10}. At redshifts $>2.0$, a
search for narrow-line objects in the SDSS-I/II quasar sample
\citep{2010AJ....139.2360S} yielded a number of NLS1, but little else.

The BOSS survey goes substantially deeper than SDSS-I/II in
spectroscopy, allowing 
us to resume the search for Type II quasars at high redshifts based on
rest-frame ultraviolet spectra. 
In this paper we adopt a composite approach based both on the 
properties of emission lines and on the properties of the continuum. 
For our parent sample, we selected all BOSS objects in DR9 with both
Ly$\alpha$ and \civ\ emission line measurements (given the BOSS
wavelength coverage, this corresponds to 
redshifts of $z \gtrsim 2.0$) and a reliable pipeline fit (i.e., the
flag {\tt ZWARNING} = 0; see the discussion in \citealt{Bolton12}).
A total of 79,505 objects satisfied these criteria.  Objects with FWHM{$ <
2000$ \kms} are quite rare, representing only 3.7\% of the total in the
BOSS sample.  These numbers are summarized in Table~\ref{tab:numbers}.

\begin{table}
\caption{Selection of Type II Candidates}
\label{tab:numbers}
\begin{center}
\begin{tabular}{lr}
\hline
Sample & No.\ of Objects\\
\hline
All BOSS DR9 quasars &102100\\
BOSS DR9 quasars with \lya, \civ\ measured &79505\\
FWHM$_{\rm \civ} < 2000\,\kms$&2494\\
\quad Class A&145\\
\quad Class B&307\\
\hline
\end{tabular}
\end{center}
\end{table}

\begin{figure}
\begin{center}
\includegraphics[width = 12cm,angle=270]{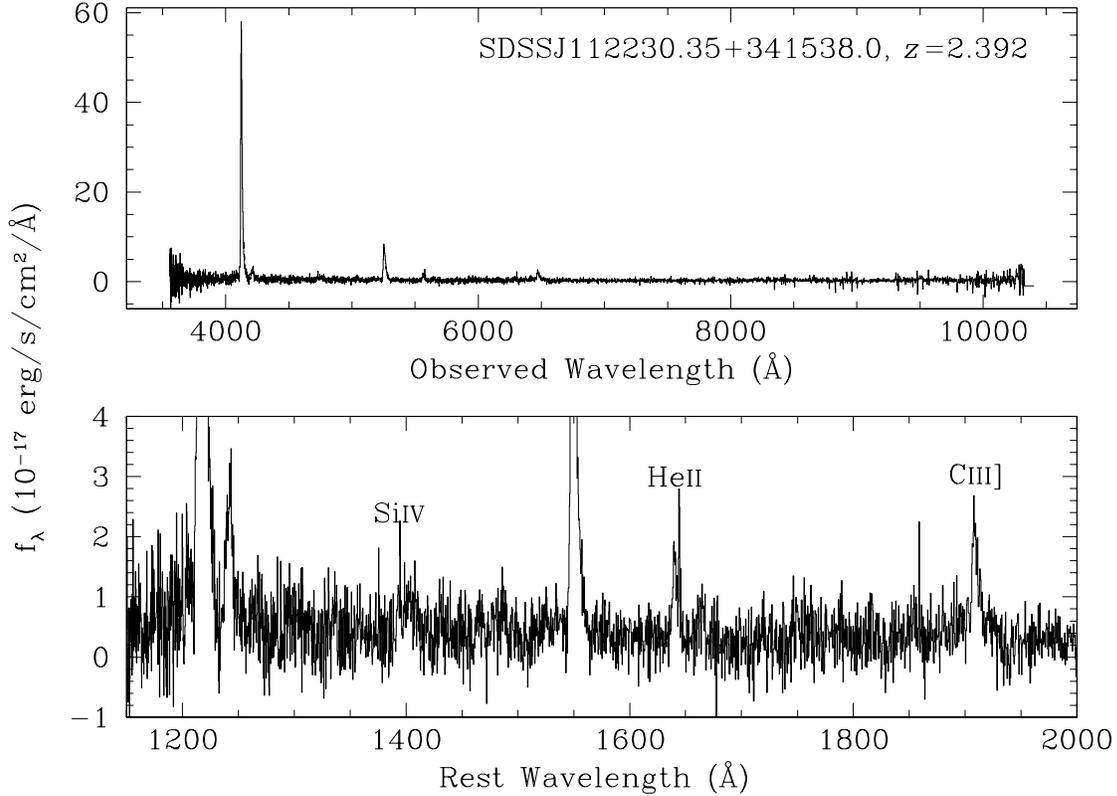}
\end{center}
\caption{BOSS spectrum of a Type II quasar candidate at $z \sim
2.4$; this is a typical object in Class A.
  Note the strong narrow emission lines (Ly$\alpha$ and \civ\ have
  FWHM of 1000 \kms\ and 1200 \kms, respectively) and weak
  continuum.  The upper panel shows the full spectrum in observed wavelengths, while the lower
  panel expands the horizontal and vertical scales and plots rest-frame wavelengths,
  with emission lines identified. In this and subsequent figures, the
  spectra have not been smoothed.}
\label{fig:quasar}
\end{figure}

\begin{figure}
\begin{center}
\includegraphics[width=12cm]{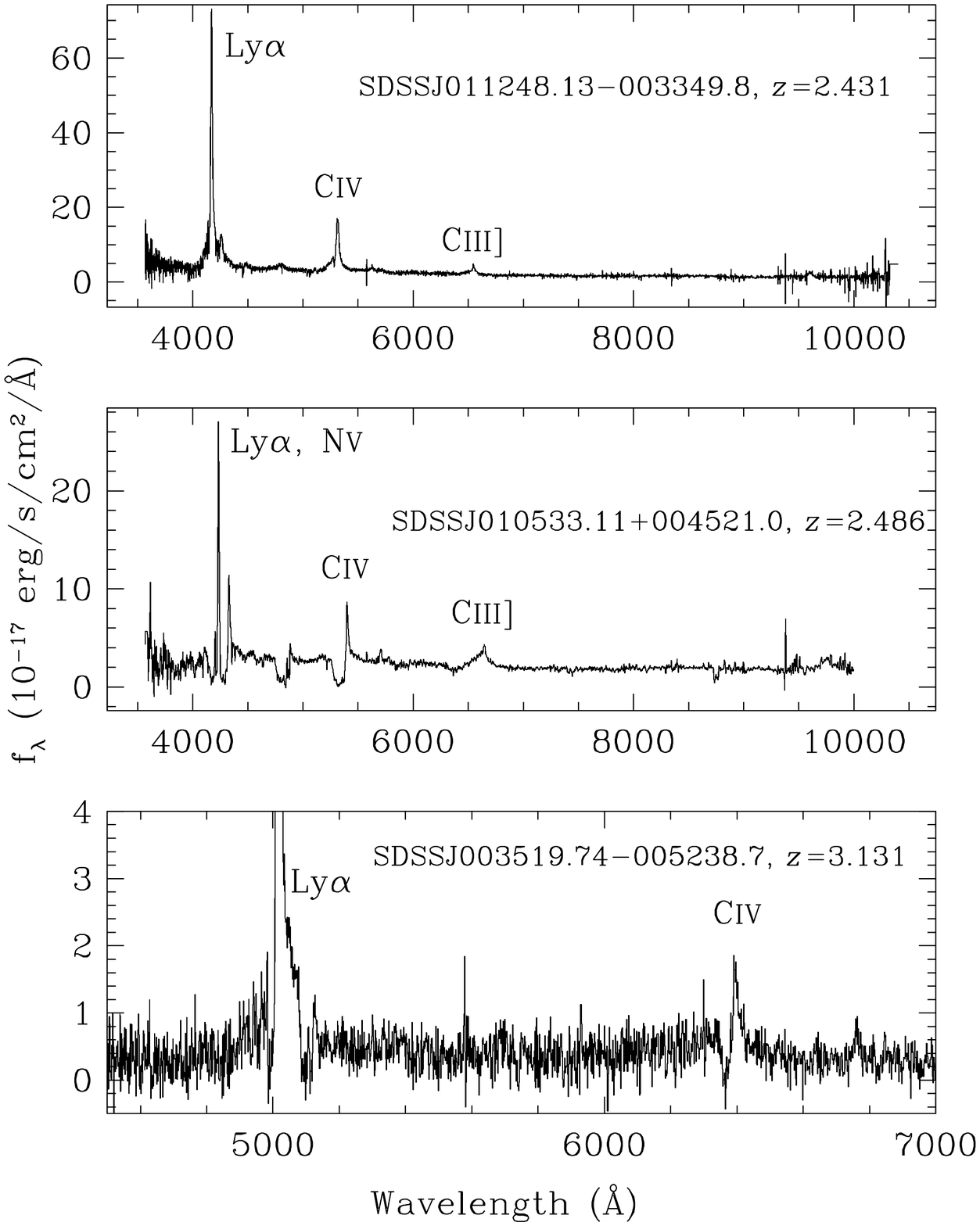}
\end{center}
\caption{
{\it Upper}: A narrow-line BOSS object that we classified as a NLS1;
note the strong blue continuum. The \civ\ and Ly$\alpha$ lines have FWHM of
  1680 and 1960 \kms, respectively as measured by the SDSS pipeline. 
  {\it Middle}: Spectrum of a
  Broad Absorption Line (BAL) quasar identified in 
  our sample of candidates.  The \civ\ emission line is truncated by the
  extensive blueward absorption; similar absorption troughs are seen
  in \siiv, \nv, and (to a lesser extent) in \ciii.  The \civ\
  emission line has a FHWM from the Gaussian fit of 900 \kms;
  Ly$\alpha$ is significantly broader, at 1900 \kms.
  {\it Lower}: An object from our Class B sample; notice the expanded
  axes.  \civ\ is strongly truncated by narrow absorption. }
\label{fig:nls1_bal}
\end{figure}

From this sample, we selected only those objects with 5$\sigma$
detections in both Ly$\alpha$ and \civ\ (thus we are insensitive to
objects that emit only in Ly$\alpha$; see, for example,
\citealt{2004AJ....127.3146H}), and restricted ourselves to
those objects in which both lines have FWHM${} < 2000$ \kms.  The
widths of the two lines are correlated, although the correlation is
weak, due in part to absorption features (\S~\ref{sec:multipeak}) and limited S/N. We then
visually inspected the spectra of the 2494 remaining candidates.  The
best Type II quasar candidates have narrow but strong emission lines
and an extremely weak, flat continuum (see Figure \ref{fig:quasar}).
There were two main astrophysical contaminants in our original sample:
high-redshift analogs to NLS1 galaxies, and broad absorption line
(BAL) quasars whose emission lines are mostly absorbed away, leaving
only a narrow component in emission (Figure~\ref{fig:nls1_bal}).  The rest-frame ultraviolet
spectra of NLS1s \citep{ConstantinShields03} show narrow permitted
lines, but possess a strong blue continuum and emission from \feii\
complexes that are not seen in obscured quasars; see for example the
upper panel of Figure~\ref{fig:nls1_bal}. An example of a BAL quasar
whose emission line widths met our initial criteria is found in the
middle panel of Figure~\ref{fig:nls1_bal}; the \civ\ line is clearly
truncated by the associated absorption.

\begin{figure}
\begin{center}
\includegraphics[width = 14cm]{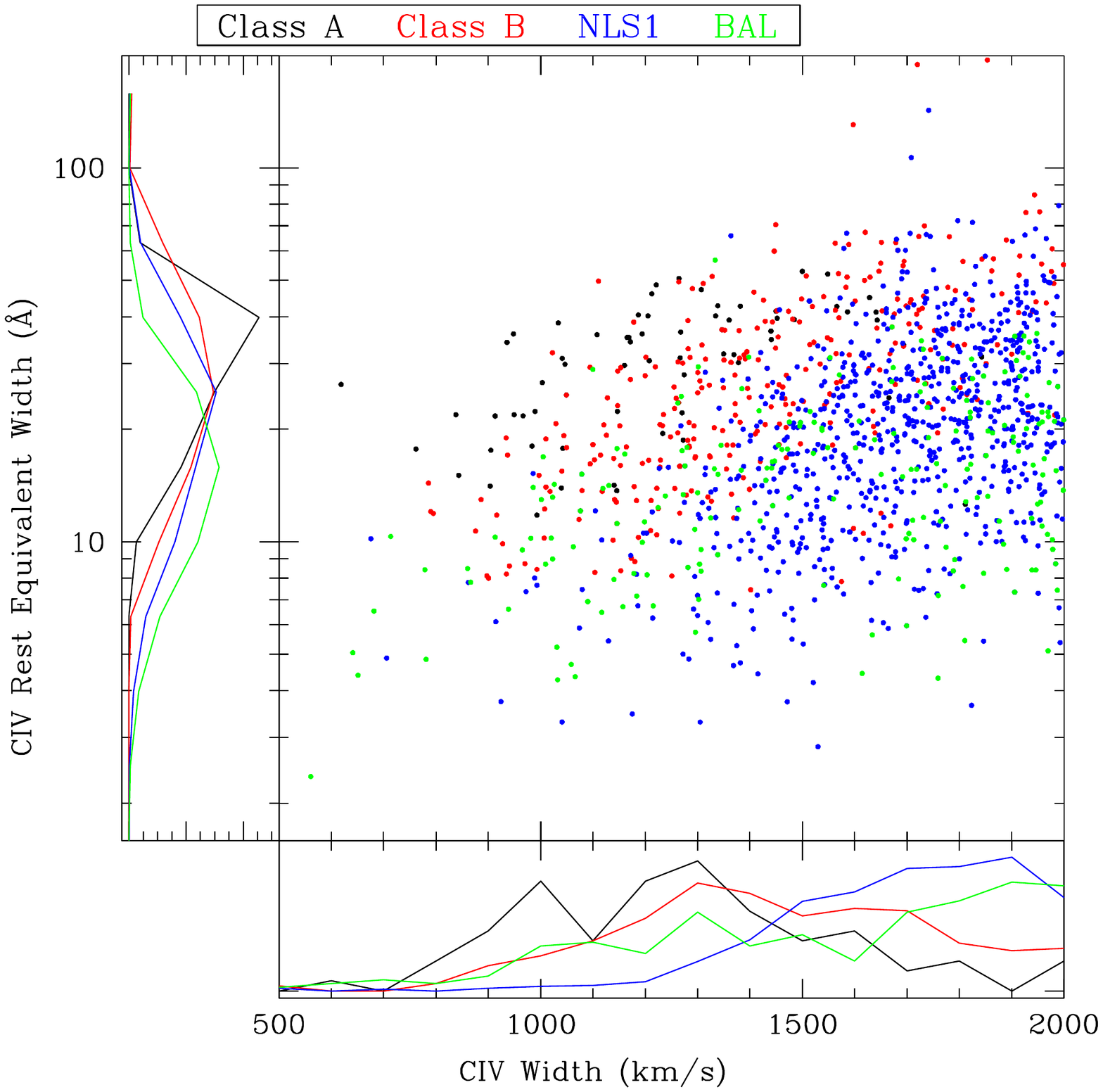}
\end{center}
\caption{The FWHM and rest-frame equivalent width (EW) of the \civ\
$\lambda$1549\AA\ line for all 2494
objects in DR9 with FWHM $< 2000\,$\kms\ and 5$\sigma$ detection of both \lya\ and \civ\
emission lines.  Type II quasar candidates are shown with class A objects in black and
class B objects in red (see the discussion in \S~\ref{sec:optical}), NLS1s are shown in blue 
and BALs in green.  The distributions of  each quantity (normalized to the same integral) 
are plotted on the sides.  The class A and class B objects tend to
have considerably lower FWHM and higher equivalent width than the NLS1 and the
  BALs.}
\label{fig:FWHM_dist2}
\end{figure}

We classified our Type II quasar candidates, after removing the BAL
quasars and obvious NLS1, in two categories: those
that showed all the qualities of an obscured quasar (narrow emission
lines, no associated absorption, weak continuum); hereafter Class A,
and those with one or more characteristics of unobscured objects (a
broad component to an emission line, an absorption feature, or a
strong blue continuum); hereafter Class B.  Figure \ref{fig:quasar}
presents an example of an object from Class A, while the lower panel
of Figure~\ref{fig:nls1_bal} shows the spectrum of a Class B
object. The Class B object has very narrow emission lines and weak
continuum, but strong narrow absorption blueward of \civ.  These
classifications are S/N-dependent and are somewhat subjective, and
there are no doubt objects we have placed in each category that belong
in the other (see Figure~\ref{fig:FWHM_dist2} and the discussion in
\S~\ref{ssec:composite_spectrum}).

Are our Type II candidate objects truly obscured quasars, in the
sense of having bolometric luminosities much larger than inferred from
the optical data?  Our objects have line widths up to 2000 \kms, while
low-redshift Seyfert II galaxies typically have line widths less than
1200 \kms\ \citep{2005AJ....129.1783H}.  However, outflows in the
narrow-line region may give significantly larger widths in more
luminous objects, with spatial extents as large as 10 kpc \citep[e.g.,][]{2012ApJ...746...86G,Liu13}.  We will
see in what follows that the obscured nature of our objects remains
unclear, and it is possible that our sample is somewhat heterogeneous,
with more than one population contributing.  With all this in mind, we
consistently refer to our objects as Type II {\em candidates}.  

Figure~\ref{fig:FWHM_dist2} shows the correlation between the
rest-frame equivalent width and FWHM of the \civ\ line for each
sample.  The distribution of objects identified as Type II quasar
candidates peaks at much smaller values of \civ\ FWHM than either
contaminating sample, but there is significant overlap among the three
distributions.  An initial cut at a smaller \civ\ FWHM would have
increased our fraction of identified Type II quasar candidates but
would have also excluded many strong candidates.

The \citet{Bolton12} pipeline redshifts are determined by fits to a
linear combination of templates that do not include narrow-line
quasars.  The asymmetries seen in Ly$\alpha$ due to the onset of the
Ly$\alpha$ forest, and especially in \civ\ due to winds
\citep{Richards11} in broad-line objects are not present for the
objects in our sample, meaning that the redshifts based on these
templates are often biased high by of order 0.005.  
The DR9 quasar catalog \citep{Paris12} includes redshifts measured directly from a
Gaussian fit to the \civ\ line, which usually lines up well with the
redshift measured from Ly$\alpha$ and fainter lines such as \ciii\ in our
narrow-line objects.  We thus use the \civ\ redshift when available in
the DR9 quasar catalog, and a redshift based on
visual inspection in the rare cases where it is not.

\section{Properties of Type II Quasar Candidates: SDSS Data}
\label{sec:optical}

\begin{figure}
\begin{center}
\includegraphics[width=12cm]{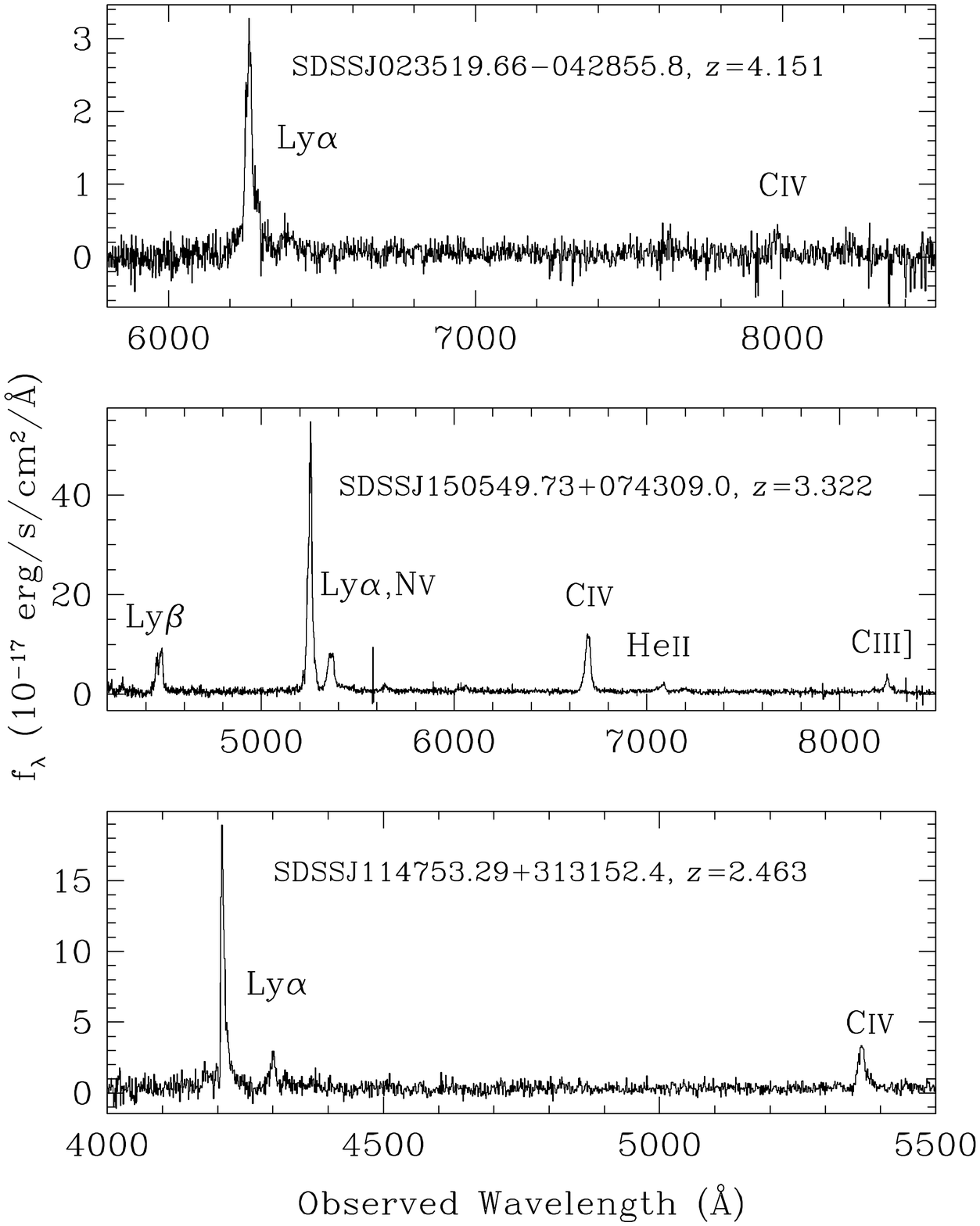}
\end{center}
\caption{The BOSS spectra of some of the notable objects in the Class A
sample of Type II quasar candidates.  {\it Top:} One of the
highest-redshift objects in our sample.  \civ\ is barely detected. 
{\it Middle:} One of the objects with highest emission-line flux. {\it
Bottom:} This object has among the narrowest emission lines in the
sample (540 \kms\ in Ly$\alpha$ and 950 \kms\ in \civ).  It also has the reddest
continuum in the sample.}
\label{fig:classA_examples}
\end{figure}

Figure~\ref{fig:classA_examples} shows spectra of further examples of objects in
our class A sample, including one of the highest redshift objects in
the sample, and the objects with the 
strongest and narrowest emission lines. 

\begin{figure}
\begin{center}
\includegraphics[width = 14cm]{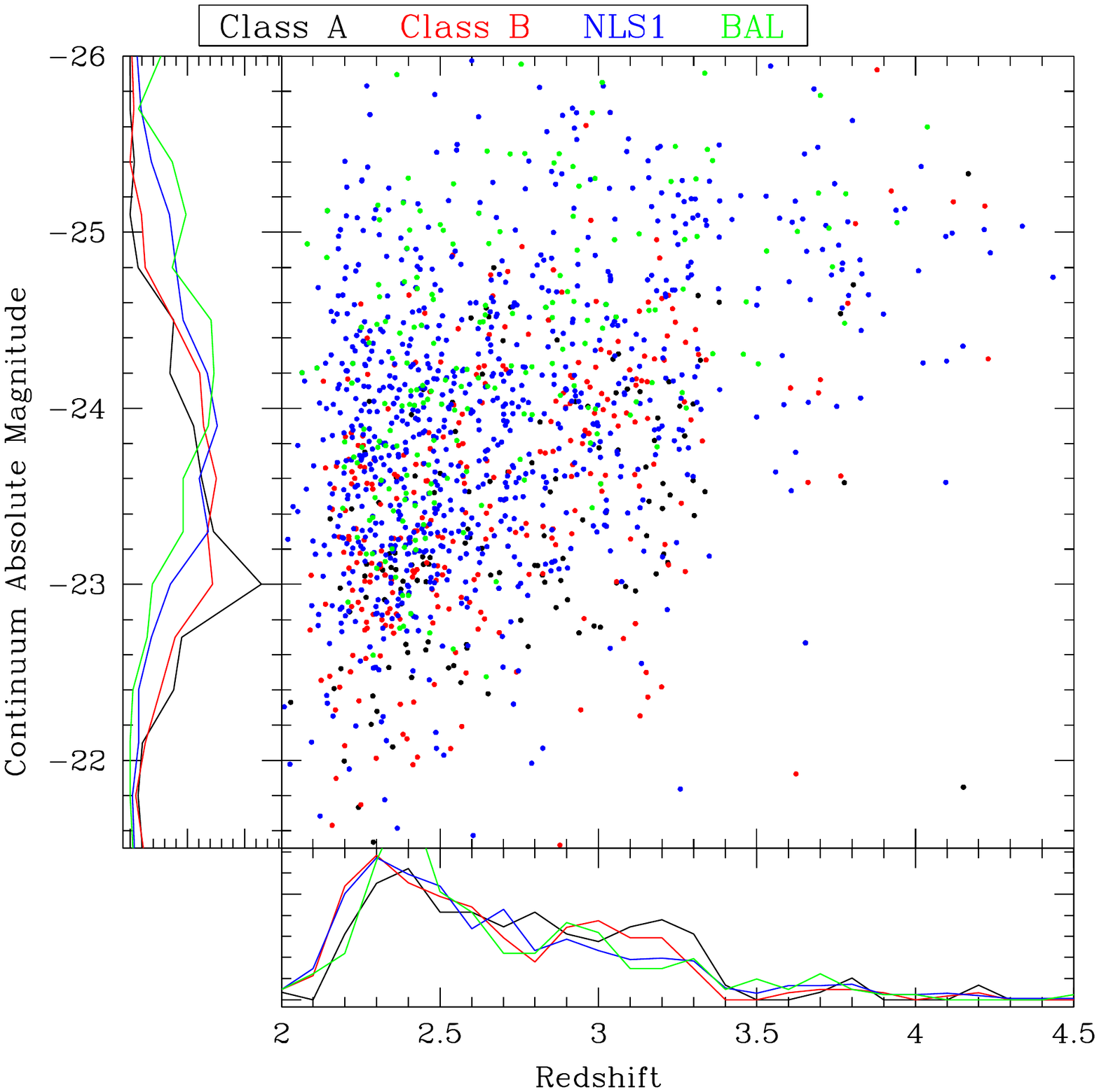}
\end{center}
\caption{Two-dimensional distribution of redshift and continuum 
absolute magnitude (calculated at a rest wavelength of $\lambda = 1450$~\AA)
for all Type II quasar candidates in our sample.  The color-coding is
the same as in Figure~\ref{fig:FWHM_dist2}.  Normalized distributions of each
individual quantity are shown on the side.  The continuum luminosities
are measured from the spectra; uncertainties in the spectrophotometric
calibration are as large as 40\%.  The Type II quasar candidate continuum
absolute magnitude distribution indicates that both the class A objects
and the class B objects tend to be fainter than the other two samples, as we would expect 
for obscured objects.  The redshift distributions of the samples are
similar to one another, and to the parent DR9 quasar sample overall.}
\label{fig:luminosity_dist}
\end{figure}

Our sample includes 452 objects that are strong or possible Type II
quasars.  We list the 145 Class A candidates in Table~\ref{tab:classA}
and the 307 Class B candidates in Table~\ref{tab:classB}. Coordinates
are given in J2000. The median
FWHM of the \civ\ line among the Class A candidates is 1260 \kms.  The
line parameters are as measured by the BOSS pipeline \citep{Bolton12};
null values correspond to problems with the measurement.  The
mean redshift of these two samples is 2.70, with a redshift range from
2.03 to 4.23 (see Figure
\ref{fig:luminosity_dist}).   In what follows, we concentrate on the class A sample.  

{\tiny
\begin{table}
\caption{Table of 145 class A candidate Type II quasars}
\label{tab:classA}
\begin{tabular}{|c|c|r|c|c|c|c|c|c|}
\hline
SDSS J2000&ra&dec&redshift&\civ\ FWHM&\civ\ REW&\civ\ flux&\lya\
flux\\
Name&deg&deg&&\kms&\AA&\ergscm&\ergscm\\
\hline
\hline
SDSSJ001040.82+004550.5&2.67008&0.7640&2.7148&1507$\pm$59&43.9$\pm$0.0&73.3$\pm$2.8&248$\pm$2.9\\
SDSSJ001738.55$-$011838.7&4.41064&$-$1.3108&3.2260&768$\pm$52&25.2$\pm$0.5&21.3$\pm$1.4&101.3$\pm$2.3\\
SDSSJ001814.72+023258.8&4.56134&2.5497&2.9024&874$\pm$97&19.1$\pm$0.8&16.67$\pm$1.9&143.6$\pm$3.0\\
SDSSJ003605.26+001618.7&9.02193&0.2719&2.9503&1773$\pm$294&12.3$\pm$1.5&18.7$\pm$3.0&76.2$\pm$2.4\\
SDSSJ004423.20+035715.5&11.09665&3.9543&2.2213&1182$\pm$47&null&74.3$\pm$3.0&416.8$\pm$7.8\\
SDSSJ004600.48+000543.6&11.50199&0.0955&2.4560&1286$\pm$61&28.1$\pm$0.5&50.3$\pm$2.5&225.2$\pm$3.9\\
\hline
\end{tabular}

Only a portion of this table is shown here to demonstrate its form and
content. A machine-readable version of the
full table will be published online.  
\end{table}}

{\tiny
\begin{table}
\caption{Table of 307 class B candidate Type II quasars}
\label{tab:classB}
\begin{tabular}{|c|c|r|c|c|c|c|c|c|}
\hline
SDSS J2000&ra&dec&redshift&\civ FWHM&\civ REW&\civ\ flux&\lya\ flux\\
Name&deg&deg&&\kms&\AA&\ergscm&\ergscm\\
\hline
\hline
SDSSJ001008.02+000317.5&2.53341&0.05485&2.2918&1655$\pm$21&null&219.4$\pm$2.8&477.0$\pm$4.7\\
SDSSJ001142.42$-$000845.7&2.92675&$-$0.14602&2.3146&1154$\pm$64&11.3$\pm$0.4&39.0$\pm$2.0&135.5$\pm$3.3\\
SDSSJ001344.04+011456.0&3.43349&1.24889&2.2250&1590$\pm$42&null&95.9$\pm$2.4&237.2$\pm$4.4\\
SDSSJ001922.82$-$004938.2&4.84507&$-$0.82727&3.3060&968$\pm$68&8.7$\pm$0.4&28.6$\pm$2.0&218.3$\pm$3.1\\
SDSSJ003519.74$-$005238.7&8.83224&$-$0.87742&3.1290&1086$\pm$85&null&25.5$\pm$1.9&166.7$\pm$2.5\\
SDSSJ003809.47+031634.1&9.53944&3.27615&2.4644&1173$\pm$118&8.8$\pm$0.7&34.0$\pm$3.5&187.9$\pm$5.2\\
SDSSJ005018.67+050132.5&12.5778&5.02571&2.9370&1149$\pm$75&12.8$\pm$0.6&46.0$\pm$3.2&214.7$\pm$3.2\\
\hline
\end{tabular}

Only a portion of this table is shown here to demonstrate its form and
content. A machine-readable version of the
full table will be published online.  
\end{table}}

\begin{figure}
\begin{center}
\includegraphics[width=12cm]{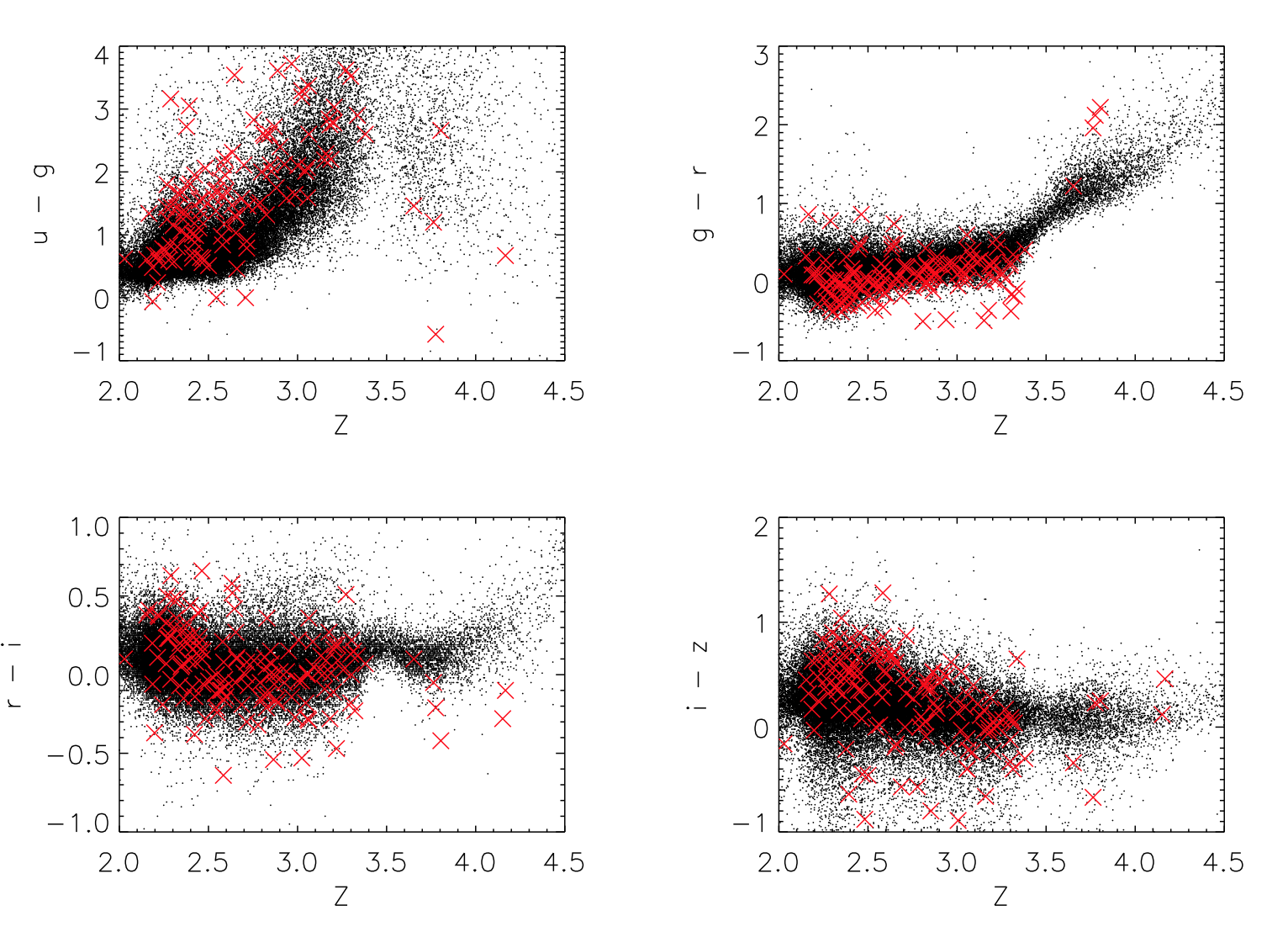}
\end{center}
\caption{SDSS colors as a function of redshift for all BOSS quasars
  in DR9 (black dots), and the 145 Class A Type II quasar candidates
  (red crosses).  A K-S comparison of the $g-r$ color distribution of the $z <
  3.5$ sample with a subsample of Type I quasars matched in $i$-band
  magnitude shows that both populations are drawn from the same distribution with
  84\% confidence.  This result suggests that the continuum is more
  quasar-like than galaxy-like.}
\label{fig:color_redshift}
\end{figure}

Figure \ref{fig:luminosity_dist} shows the continuum ($\lambda =
1450$\AA) luminosity density distribution for our candidates, as well
as for those objects we identified as NLS1s and BALs.  These continuum
luminosities, which are measured from the BOSS spectra, are 
significantly larger than would be expected from the host galaxy of
the quasars; the most luminous non-active galaxies at these redshifts
have absolute ultraviolet continuum magnitudes $-22.5$ \citep{Shapley11}.
For comparison, the Type I quasar luminosity function has been
measured down to $M=-25$ at these redshifts from the BOSS survey
\citep{Ross12b}.  The high luminosities of our objects, 
together with the quasar-like broad-band colors (see discussion in section 
\ref{ssec:continuum_slope} and Figure \ref{fig:color_redshift}), suggest that the
quasar continuum of these objects is not completely extincted.  These
objects may have only a modest dust column, or perhaps the observed
continuum is due to light from the central engine that is reflected
into our line of sight by dust in the quasar host galaxy, as is found
in lower-redshift Type II quasars \citep{2006AJ....132.1496Z}.  A
third possibility is that the extinction is patchy, with unextincted
light showing through holes in the obscuring dust.  
  
\subsection{Broadband Colors}
\label{ssec:continuum_slope}

Almost all the objects in both Class A and Class B were selected for
spectroscopy as quasar candidates, using the algorithms described in
\citet{2012ApJS..199....3R}.  There are four principal algorithms
used, but 60\% of the objects in our sample were selected using the likelihood algorithm
of \citet{2011ApJ...743..125K} and the probabilistic algorithm of
\citet{2011ApJ...729..141B}, both of which use the distribution of
objects in color space, incorporating information on photometric
errors.  The broad-band colors of ordinary quasars as a function of
redshift are a reflection of the blue continuum, absorption by the
Ly$\alpha$ forest and Lyman-limit systems, and, to a lesser extent,
emission lines in the various bands \citep{Fan99,2003AJ....126.1131R}.
Understanding the demographics of obscured quasars
will require quantifying the biases that the target selection
algorithms impose on the sample.  

SDSS imaging is carried out in five bands, $ugriz$ \citep{1996AJ....111.1748F}.
Figure~\ref{fig:color_redshift} shows the dependence of color on
redshift, both for our sample of Type II quasars, and the full sample
of DR9 quasars \citep{Paris12}.  The mean colors of the two groups are roughly the same,
suggesting that the continua of objects in our sample are quasar-like.
Of course, these objects were selected by their broad-band colors, and
it perhaps not surprising that their colors are similar to those of
other objects selected in the same way.  The rest-frame equivalent
widths of \lya, typically the strongest line in these objects, can be
as large as 1000\AA, at which point the line starts significantly affecting
the broad-band colors.   

As the
Ly$\alpha$ break enters the $g$ band at $z \sim 3.5$, the $g-r$ colors
quickly redden with redshift.  To quantify the similarity of the
colors to unobscured quasars, we selected a subsample of DR9 quasars
with similar $i$-band magnitude distribution as our Type II quasar
candidates, so that they have similar photometric errors.  Only
objects from our sample and the DR9 catalog with a redshift of $z
\leq 3.5$ were used in the comparison, so the color was close to
constant over the whole sample.  A Kolmogorov-Smirnov (K-S) test of
the $g-r$ color for each sample shows that the two distributions are
the same at the 84\% confidence level, thus the colors of our objects
are consistent with having been drawn from the distribution of
unobscured quasars.

\subsection{Composite Spectrum}
\label{ssec:composite_spectrum}

\begin{figure}
\begin{center}
\includegraphics[width = 12cm]{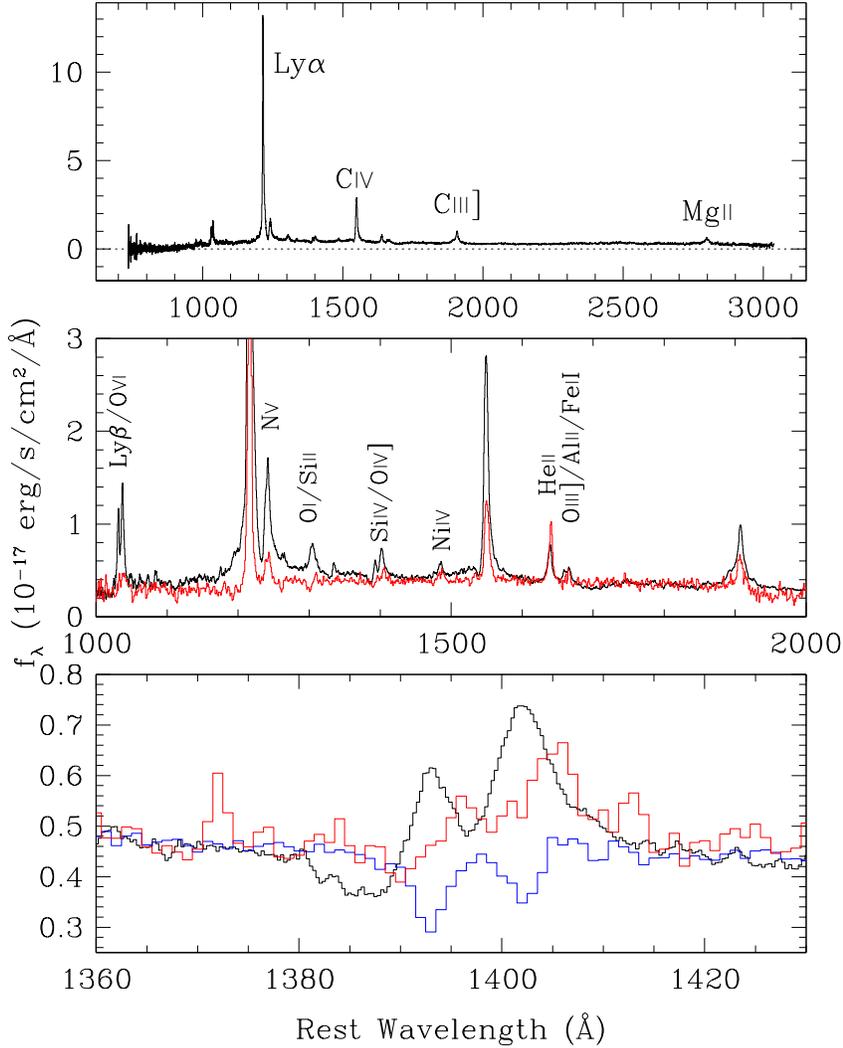}
\end{center}
\caption{{\it Upper:} Composite spectrum of all 145 Class A Type II quasar
  candidates; the dotted line is drawn at zero flux.  {\it Middle:} An expanded view of the composite, to show the
  weaker emission lines. Note the absorption blueward of \civ\ and
  \siiv/\oiv.  Superposed in red is the normalized composite of the 
  \citet{2011ApJ...733...31H} sample of high-redshift AGN, normalized
  to roughly match the continuum level at 1700\AA.  The
  continuum slopes of the two composites are quite similar.  The
  Ly$\alpha$ line in the \citet{2011ApJ...733...31H} composite peaks at a flux
  density level half of that in the Class A composite.  {\it Lower:}
  An expansion of the composite in the region around the \siiv\
  doublet.  The black line is our composite (lightly smoothed).  The two
  emission features are the \siiv\ doublet at 1393.76~\AA\ and
  1402.77~\AA, respectively; note the absorption trough blueward of it.
   The blue curve shows the composite LBG spectrum of
   \citet[][normalized to the same continuum]{Shapley03}; the
   absorption features precisely align with 
   the emission lines in our composite.  The red curve shows the
   composite spectrum of Type II objects from
   \citet{2011ApJ...733...31H}.  Intriguingly, the \siiv\ lines are
   redshifted relative to rest frame by about 800 \kms, perhaps due to
   systematic absorption on their blue wing. }
\label{fig:coadd}
\end{figure}

The upper and middle panels of Figure~\ref{fig:coadd} show the
arithmetic average spectrum
of our 145 Class A Type II quasar candidates.  This average is calculated by
shifting all spectra to the rest frame, and inverse-variance weighting
the spectra at a given wavelength, using errors estimated by the SDSS
spectroscopic pipeline.  

We used the program {\tt MPFIT} \citep{2009ASPC..411..251M} in IDL to
fit one or two Gaussian components and a local continuum around each
emission line in the composite spectra.  \civ\ was fit with two
Gaussians corresponding to a broad and narrow line component, after
masking any associated absorption blueward of the emission line.
Other than Ly$\alpha$, the redshifts and widths of emission lines were
constrained to be the same as that of \civ. The
\lya\, \nv, \heii\ and
\ciii\ lines were also each fit with two Gaussian components, 
while all other emission lines were fit with only a single Gaussian
component.  \lya\ was fit simultaneously with \nv, \siiv\ was fit
simultaneously with 
\oiv, and \ciii\ was fit simultaneously with both \siiii\ and \aliii.
The resulting fluxes are listed in Table~\ref{tab:flux}, and the
measured emission line widths are listed in Table~\ref{tab:width}.
The continuum level in the composite is constant at about $0.4 \times
10^{-17}$ erg s$^{-1}$ cm$^{-2}$ \AA$^{-1}$, so rest-frame equivalent widths in \AA ngstroms are roughly
the listed flux value, times 2.5.

Interestingly, the \lya\ and \ciii\ emission lines in our composite
Type II candidate spectrum show a broad base ($\rm FWHM >3000$ \kms;
Table~\ref{tab:width}), as does \civ\ after
masking absorption blueward of the line.  The flux in this broad
component is comparable to the narrow component, as
Table~\ref{tab:flux} makes clear.  This absorption and broad base were
among the qualities that caused us to classify some of our candidates
as Class B; at higher S/N, it is likely that many of the individual
Class A objects would show broad components or associated \civ\
absorption.  The composite spectrum of our NLS1s shows significantly
stronger broad bases on all of the emission lines as well as a
significantly higher continuum flux relative to the emission lines. In
addition, absorption blueward of \civ\ is particularly strong.

\begin{table}
\caption{Measured emission line fluxes from the composite spectra of our Class A, 
Class B and NLS1 samples.  For strong emission lines where two components were fit
we include values for both a narrow and broad component; see
Table~\ref{tab:width}).}
\begin{center}
\begin{tabular}{|l|l|r|r|r|}
\hline
\hline
Emission Line&Wavelength& Class A  Flux & Class B  Flux & NLS1 Flux \\
&(\AA)&(\ergscm)&(\ergscm)&(\ergscm)\\
\hline
\hline
\lya\ (narrow)&1216&35.1&28.5&26.3\\
\lya\ (broad)&1216&37.9&28.6&34.8\\
\nv\ (narrow)& 1240&3.4&3.2&4.4\\
\nv\ (broad)&1240&9.7&12.3&22.8\\
\oi\ &1305&1.8&1.2&1.3\\
\cii\ &1337&0.24&0.22&0.24\\
\siiv\ &1397&0.34&0.4&1.0\\
\oiv\ &1402&1.9&1.4&1.9\\
\niv\ &1486&0.76&0.25&0.13\\
\civ\ (narrow)&1549&10.5&9.3&10.8\\
\civ\ (broad)&1549&12.3&13.1&18.5\\
\heii\ (narrow)&1638&1.8&1.2&1.4\\
\heii\  (broad)&1638&0.8&1.6&2.5\\
\oiii\ + \alii\ + \feii\ &1665&0.89&0.66&0.67\\
\aliii\ &1857&0.24&0.33&0.60\\
\siiii\ + \feiii\ &1892&0.73&0.25&0.49\\
\ciii\ + \feiii\ (narrow)&1906&1.4&1.8&2.1\\
\ciii\ + \feiii\ (broad)&1906&6.7&6.4&9.7\\
\hline
\end{tabular}
\end{center}
\label{tab:flux}
\end{table}%

\begin{table}
\caption{Measured emission line widths (FWHM) from the composite spectra. We include fits 
to two Gaussians, a narrow and broad component.}
\begin{center}
\begin{tabular}{|l|l|l|l|l|}
\hline
\hline
Emission Line &Wavelength& Class A Width& Class B Width& NLS1 Width\\
&(\AA)&(\kms)&(\kms)&(\kms)\\
\hline
\hline
\lya\ (narrow)&1216&900&1040&1150\\
\lya\ (broad)&1216&3130&3740&4140\\
\civ\ (narrow)&1549&1060&1220&1370\\
\civ\ (broad)&1549&3630&6170&7210\\
\hline
\end{tabular}
\end{center}
\label{tab:width}
\end{table}%

The middle panel of Figure~\ref{fig:coadd} expands the vertical and horizontal scales to make weaker
emission lines visible.  We detect features that are
usually blended in quasar spectra (\citealt{vandenberk01}; see also
the discussion in \citealt{HewettWild}), such as
the \siiv$\lambda$1393,1402\AA\ doublet 
and \ciii$\lambda1909$\AA, 
Si{$\,$\scriptsize III]}$\lambda 1892$\AA\ and Al{$\,$\scriptsize III}$\lambda
  1857$\AA, are resolved.   The 
  N{$\,$\scriptsize IV}$\lambda$1486\AA\ line, which is only rarely seen in Type
  I AGN \citep{2004AJ....128..561B,2008ApJ...679..962J}, but is seen
  in high-redshift radio galaxies \citep{2008MNRAS.383...11H} and
  low-luminosity Type II AGN \citep{2011ApJ...733...31H} is present.  
  There is little evidence for the \feii\ emission complexes blueward
  of \mgii\ \citep{vandenberk01}.  
  The mean Type II candidate spectrum is blue in $f_\lambda$, although not
as blue as in NLS1s.  The continuum break due to the Ly$\alpha$ forest
is clearly visible.

  There are weak absorption features apparent in this composite, but
  they mostly appear blueward of emission lines, representing outflows
  from the central engine or the superposition of individual
  absorption features \citep{2008MNRAS.386.2055N}, rather than stellar or
  interstellar medium lines.  At
  high S/N, the spectra of Lyman-break galaxies (LBG) also show
  absorption lines in the rest-frame ultraviolet, as is apparent in
  the composite spectrum of \citet{Shapley03}; high-ionization
  absorption lines have also 
  been seen in high-redshift Type II AGN \citep{2002ApJ...576..653S,
  2011ApJ...733...31H}. In LBGs, resonance 
  absorption in the interstellar medium tends to be blueshifted by
  several hundred \kms, indicating strong outflows presumably powered
  by starbursts.   In our composite, the only absorption feature that
  appears to coincide with those in LBGs is \siiv$\lambda 1396$\AA, as
  shown in the lowest panel of Figure \ref{fig:coadd}; 
  the absorption is considerably blueward
  of the emission line, corresponding to a blueshift of order 2000
  \kms\ (the absorption feature blueward of \civ\ is also offset by
  $\sim$2000 \kms).  Moreover, it is unclear what is the driving source of these 
  outflows (processes in the host galaxy or the central engine) 
  and whether they are occurring on the galactic or on the circumnuclear scale.

\subsection{Comparison to Other Samples of Obscured Quasars}
\label{sec:comparison}

We now compare the spectral properties of our objects to those of
other obscured AGN samples in the literature.
\citet{2011ApJ...733...31H} have identified 33 AGN selected by their
ultraviolet emission-line properties from a spectroscopic survey of star-forming
galaxies at $z \sim 2-3$.  These objects, all of which show narrow
emission lines, are significantly fainter than ours, with typical
continuum brightnesses of $R \sim 24$.  The ultraviolet/optical SEDs
of these galaxies are well-fit by stellar population synthesis
models with no AGN component \citep{2012ApJ...760...74H}. Their composite spectrum is
shown in the middle panel of Figure~\ref{fig:coadd}, arbitrarily
normalized to have a similar continuum at 1700\AA\ as our composite, allowing an
approximate comparison of the rest-frame equivalent widths of emission lines in
the two composites.  The emission lines in the Hainline et al.\
composite show no evidence for a broad base in Ly$\alpha$ or \civ.
Unlike our Type II candidates, there is no hint of contamination from
an unobscured component.  However, objects from the Hainline et al.\ sample have 
emission line widths comparable to the
narrow component of the Class A composite.  One is also struck by the
significantly greater strength of all emission lines in the Class A
composite; Ly$\alpha$ (which goes off-scale in this expanded view)
peaks at $\sim 7 \times$ \ergscmA\ in the normalized Hainline et al.\ composite,
and 14 $\times$ \ergscmA\ in the Class A objects.  The one exception is the \heii\ line
at 1640\AA, which is comparable in strength in the two composites;
we'll discuss this line further below.

The peaks of Ly$\alpha$, \civ, and most weaker lines of our composite
align well with those of the Type II composite from Hainline et al.,
with the interesting exception of 
the \siiv\ 1394, 1403$\lambda$\AA\ doublet (lower panel of Figure~\ref{fig:coadd}) 
which is redshifted in the Hainline et al.\ composite relative to 
ours by 800 \kms. A similar redshift is present in the \oi/\siii\ 
1302\AA\ blend as well. Excess absorption due to 
outflows in Hainline et al. objects would suppress the blue wings 
of the emission features and shift them systematically to the red, 
although the amount of shift seems extreme. Moreover, the composite 
of our Class A objects also shows absorption, but nevertheless the 
emission peaks at exactly the expected wavelengths. Indeed, 
the \siiv\ lines we see line up well 
with the \siiv\  absorption features from the host galaxy in the LBG 
composite of \citet{Shapley03}. 

 The Hainline et al.\ galaxies have continua presumably dominated by
 their host galaxies, which makes it intriguing that the ultraviolet spectral
 shape of the composite is similar to that of the Class A composite.
 As Figure~\ref{fig:color_redshift} showed, the broad-band colors of
 our objects are similar to those of unobscured quasars.
 The continuum of the \citet{Shapley03} LBG composite
 is significantly bluer than either of these samples, reflecting
 on-going star formation in this ultraviolet-selected sample.  Similarly, the
 star-forming galaxy at $z=2.3$ studied by \citet{2010ApJ...719.1168E}
 is much bluer than our composite.  \citet{2012ApJ...760...74H} point
 out that AGN spectral lines are strong in LBG spectra in the
 most luminous objects, which tend to be red; this is why
 the objects in their sample are redder than typical LBGs.  In this
 interpretation, the similarity in continuum shape between our
 composite and that of Hainline et al.\ is fortuitous. 

\begin{figure}
\begin{center}
\includegraphics[width = 12cm]{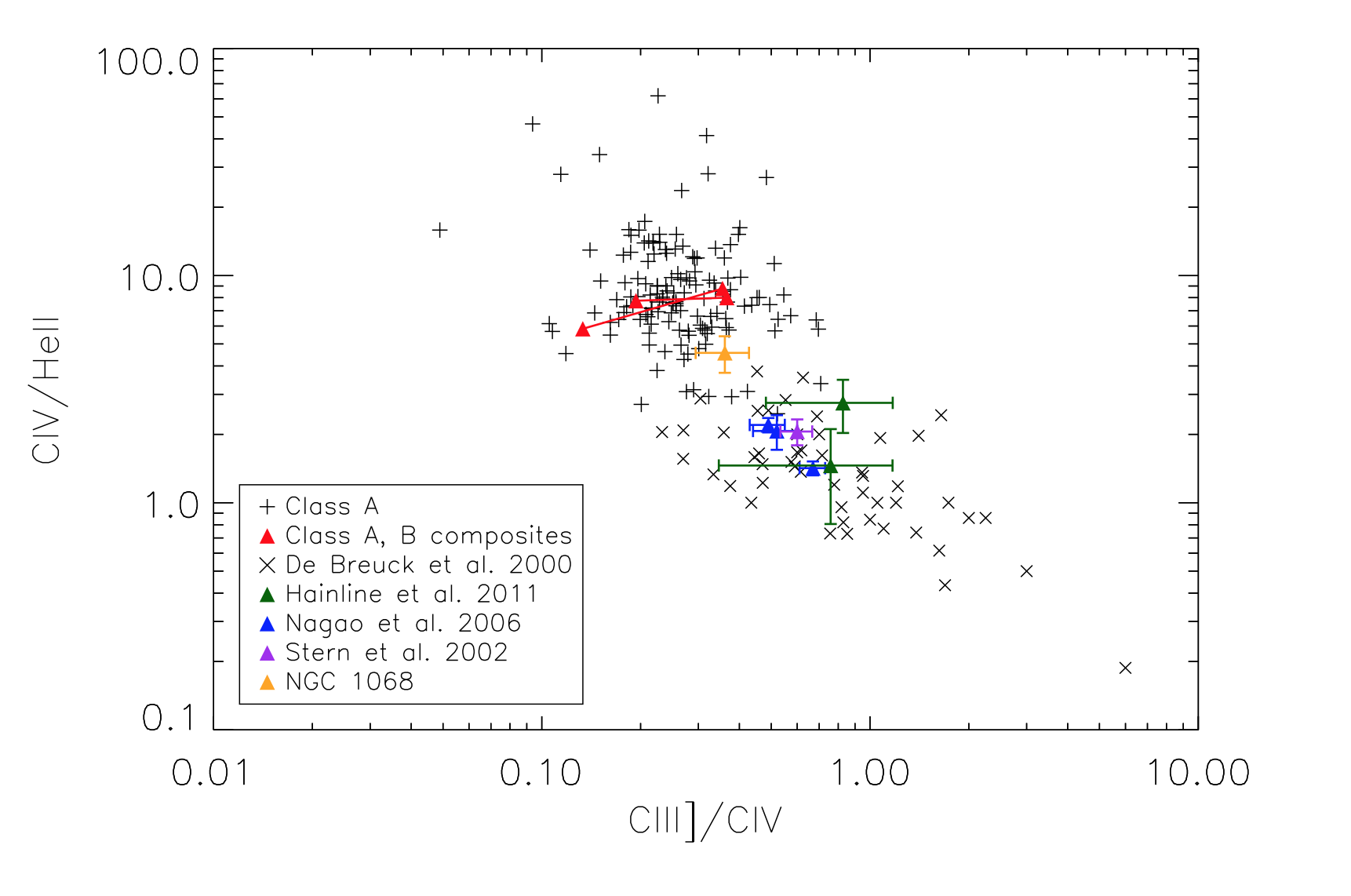}
\end{center}
\caption{Comparison of \civ/\heii\ and \ciii/\civ\
    for several samples of narrow-line AGN.  The plus signs (`+') are
our Class A candidates, and the red triangles are from the narrow
components of the measured emission lines (left) and the
 broad+narrow emission lines (right, connected with a line) of
our class A and class B composites. The green points are composites
from the work of \citet{2011ApJ...733...31H} where they split their
population at a \lya\ EW of 63\AA.  The blue points are composites
(X-ray selected narrow line quasars, low-redshift Seyfert IIs and
high-redshift radio galaxies) from the compilation of
\citet{2006A&A...447..863N} while the x's are high-redshift radio
galaxies from \citet{2000A&A...362..519D}.  We also show the
Chandra-selected Type II quasar at $z = 3.29$ of
\citet{2002ApJ...568...71S}, and ultraviolet observations of the
low-redshift Seyfert II galaxy NGC 1068 \citep{2000ApJ...532..256K}.
}
\label{fig:compare_UVemission}
\end{figure}

 The comparison in the middle panel of Figure~\ref{fig:coadd}
indicates that the relative strengths of emission lines in our objects
and those of \citet{2011ApJ...733...31H} are quite different.
Following the emission-line diagnostic diagram of
\citet{1997A&A...323...21V}, Figure~\ref{fig:compare_UVemission} compares the ratio of the
strengths of \civ\ and \heii\ to the ratio of \ciii\ to \civ\ for our
objects and those of \citet{2011ApJ...733...31H}, the compilation of
narrow-line X-ray sources and radio galaxies of
\citet{2006A&A...447..863N}, the radio galaxies of
\citet{2000A&A...362..519D}, the X-ray-selected Type II quasar of
\citet{2002ApJ...568...71S}, and ultraviolet observations of the archetypal
Seyfert II galaxy NGC 1068 \citep{2000ApJ...532..256K}.  We also show
the results for the Class A and Class B composites (red triangles),
where we measure both the narrow components of the emission lines, and
the sum of narrow and broad components.  The collisional excitation of
\civ\ is sensitive to density and to gas temperature and thus metallicity.  \civ\ can appear 
in emission in star-forming galaxies, 
although it is often swallowed by associated absorption in that line. 
However, the amount of \civ\ emission relative to other lines is much 
smaller in starbursts than in active nuclei. \ciii/\civ\ depends on
the relative abundance of these two ionization states of carbon, 
and thus this ratio is sensitive to the ionization parameter. 
It is about 0.5 in active nuclei \citep{2002ApJ...576..653S}, whereas 
it is 3 or greater in even the most extreme starbursts 
\citep{2010ApJ...719.1168E}. Our objects fit squarely in the AGN regime;
 furthermore, we find that this ratio is considerably smaller for the 
 narrow components of both Class A and B composite spectra 
 than for the full lines (Figure~\ref{fig:compare_UVemission}).
 Similarly, \heii\  may be produced in Wolf-Rayet stars or in extreme 
 star-formation regions with high ionization parameters, a hard
 spectrum, and low density, 
 but this line has an equivalent width of no more than 2.7\AA\ in 
 starbursts \citep{2010ApJ...719.1168E}, with only a handful of 
 detections known. We detect \heii\ in many individual spectra, 
 and the equivalent width in the composite spectrum is about 6\AA. 

 The ratio of \civ\ to \heii\ is significantly higher, and the ratio
of \ciii\ to \civ\ lower, than in the other Type II samples we've discussed
(with the exception of NGC 1068).  Indeed, the narrow-line region
photoionization models of
\citet{2004ApJS..153...75G} do not predict \civ/\heii\ greater than 3
for any choice of their parameters, while the typical value for our
sample is a ratio closer to 10.  These high ratios are seen in 
broad-line regions; for example, the Type I quasar composite spectrum
of \citet{vandenberk01} has $\rm \civ/\heii\ \sim 50$, and the Seyfert
I galaxy NGC 5448 \citep{2000ApJ...536..284K} has line ratios which
place it in the middle of the cloud of Class A objects in
Figure~\ref{fig:compare_UVemission}.  So these narrow-line objects
have line ratios like that of broad-line objects, but equivalent
widths that are significantly higher.  Indeed, our composite spectrum
is like those  of the ultraviolet spectrum of the Seyfert I spectrum
NGC 5548 in its low state \citep{1998ApJ...495..718G}, which shows
narrow lines with broad bases, and emission-line ratios similar to our
objects.  An interesting possibility is that some of our objects may
have similarly been observed spectroscopically in a low state; repeat
spectroscopy and comparison of the photometry and spectrophotometry
can test this hypothesis.  

\subsection{Associated Absorption}
\label{sec:multipeak}

\begin{figure}
\begin{center}
\includegraphics[width=12cm,angle=270]{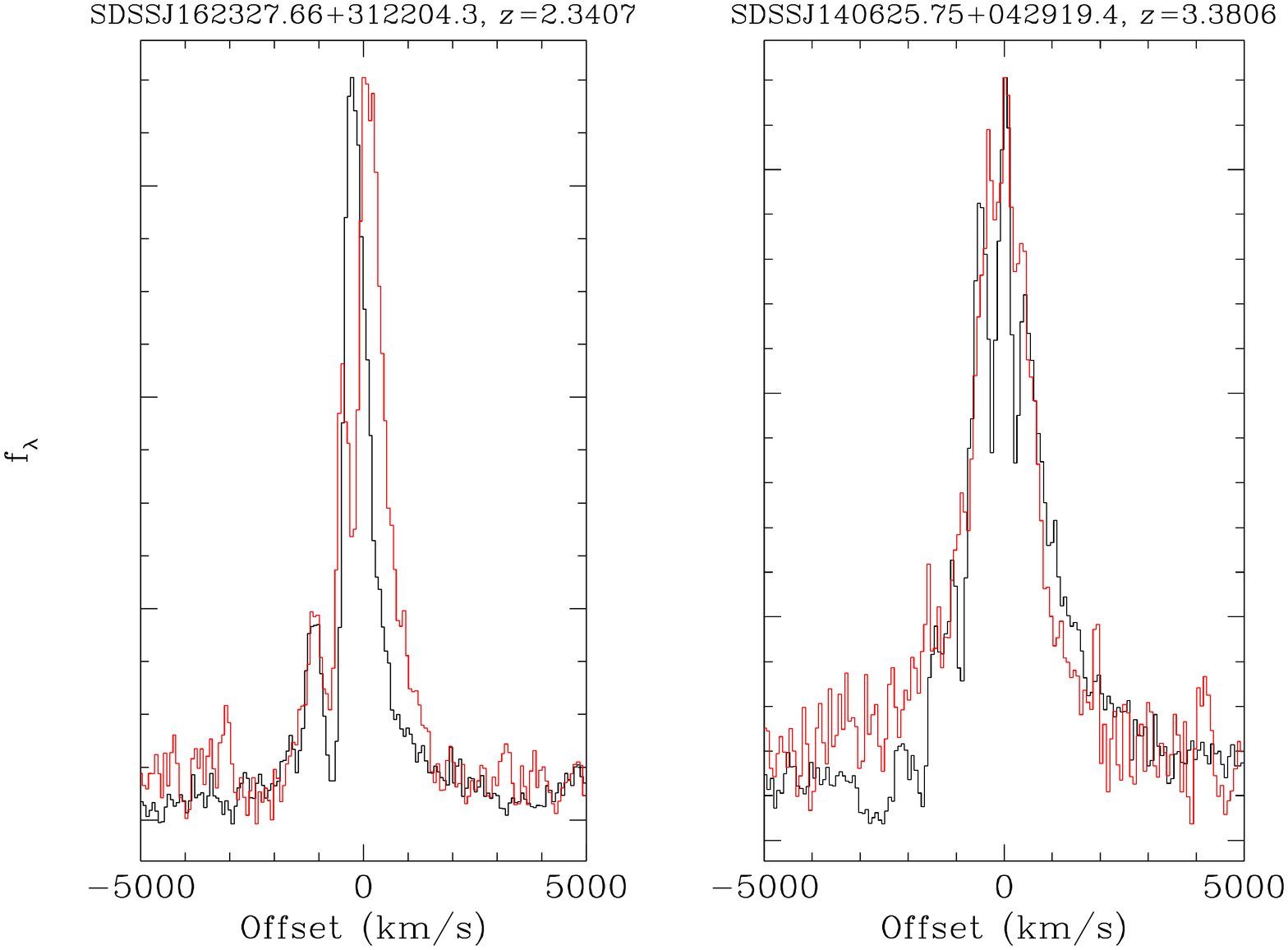}
\end{center}
\caption{Ly$\alpha$ (black) and \civ\ (red) spectra for two Class A objects,
represented in velocity space.  The two
lines have been scaled to the same maximum.  These objects have been chosen for
their prominent associated absorption in these two lines.  While the profiles of the
two lines are similar, Ly$\alpha$ shows additional absorption
features, suggestive of low-ionization gas.  Some of the structure in
the \civ\ line reflects the fact that it is a doublet.  About 12\% of
our Class A objects show such associated absorption in their BOSS
spectra.
\label{fig:velocity_multipeak}}\end{figure}

The emission line widths have been measured by fitting Gaussian
profiles to the spectra. However, the line profiles often differ significantly from
Gaussian.  A particularly dramatic example is shown in
Figure~\ref{fig:velocity_multipeak}, where the \lya\ and \civ\ lines
show multiple peaks. This is at first glance reminiscent of low-redshift
Type II objects possessing emission lines with two or more peaks, interpreted as multiple
AGN or biconical outflows \citep[e.g.,][]{Liu10a,2012ApJ...753...42C}.
However, a more likely interpretation of these systems is that they
are due to associated absorption within the host galaxy of the AGN;
associated \civ\ absorption is seen in the spectra of quasars with
velocity offsets as large as 12,000 \kms\ \citep{2008MNRAS.386.2055N}. 
Some of the absorption features in Figure~\ref{fig:velocity_multipeak}
match in Ly$\alpha$ and \civ, but inspection of the emission-line
profiles of all the objects in our sample shows many more cases in
which Ly$\alpha$ displays absorption and \civ\ does not, as expected in
sufficiently low-ionization gas, perhaps from neighboring galaxies.  About 12\% of our
Class A objects possess significant absorption in one or both of these
lines; it may be that higher S/N spectra at higher resolution would
show more.  A handful of objects show absorption in \civ\ and not
Ly$\alpha$, which may point to hot gas in which all the hydrogen is
ionized.  

\section{Sample Properties from Radio to X-Ray}
\label{sec:other_bands}

We now turn to the properties of our objects revealed by data beyond the
SDSS itself.  We match the sample with radio data from the FIRST
survey in \S~\ref{ssec:FIRST} and near-infrared data from 
the Wide-Field Infrared Survey Explorer (WISE;
\citealt{2010AJ....140.1868W}) in \S~\ref{sec:WISE}.  Several sources
have also been serendipitously detected at 24$\mu$m by the Spitzer
Space Telescope (\S~\ref{sec:spitzer}). 
We also used NASA's High Energy Astrophysics Science Archive Research
Center (HEASARC\footnote{\tt http://heasarc.gsfc.nasa.gov}) to search
for serendipitous X-ray coverage from the 
XMM-Newton, Chandra and ROSAT facilities (\S~\ref{ssec:xray}).  One of
our sources is included in the Cosmic Evolution Survey (COSMOS) field
\citep{Scoville07,2009A&A...497..635C} (\S~\ref{ssec:cosmos}).
Finally, we present optical polarization data for two of our sources
in \S~\ref{sec:polarization}.  

\subsection{Radio: Faint Images of the Radio Sky at 20cm}
\label{ssec:FIRST}

The FIRST survey \citep{1995ApJ...450..559B} covers most of the
SDSS footprint and was performed using the National Radio Astronomy
Observatory (NRAO) Very Large Array (VLA) in its B-configuration in
two channels at 1365 and 1435 MHz ($\sim$20 cm).  Images were produced
covering $1.8''$ pixel$^{-1}$ with an rms of 0.15 mJy and a resolution of $5''$.
Roughly 30\% of FIRST sources have optical counterparts in SDSS
imaging \citep{2002AJ....124.2364I}.

Radio-loud Type II quasars are well-studied at high redshift
\citep{1993ARAA..31..639M}; the typical high-redshift radio galaxy
shows strong narrow emission lines reminiscent of the objects in our
sample.  There are only 12 objects in our sample of 452 Class
A and B candidates (less than 3\%) that match within $3''$ of a FIRST
source\footnote{using the on-line catalog at {\tt
http://sundog.stsci.edu/}}; of these, only three are among the 145 Class A
candidates.  These are listed in Table~\ref{tab:radio}. However, the depth of the FIRST survey is insufficient to
designate an undetected object in our sample as radio-quiet \citep{2002AJ....124.2364I}.  Moreover, this
classification depends on the ratio of optical to radio flux, and if
the optical flux in our sources is significantly extincted, 
the optical-to-radio ratio is meaningless. 
Indeed, while 8-20$\%$ of all
Type I AGN are radio-loud \citep{2004AJ....128.1002Z}, only 1.5\% of
Type I quasars in the DR9 quasar catalog with $21.5 < i < 22$ and
redshifts between 2 and 3 are detected in the FIRST catalog.  The
radio-loud fraction of Type I quasars is lower at higher redshift
\citep{2007ApJ...656..680J}; it is possible that this also affects the
number of radio detections in our sample.  

Of the 12 matches, only one, SDSSJ081257.15+181916.8, appears double-lobed in
the FIRST images, with a separation between the lobes of roughly $10''$,
corresponding to a physical separation of about 80 kpc.  

\begin{table}
\caption{FIRST Survey radio detections of Class A and Class B
candidates.}
\begin{center}
\begin{tabular}{|l|r|r|}
\hline
\hline
Name&Peak&Integrated\\
&Flux&Flux\\
&(mJy/beam)&(mJy)\\
\hline
\hline
Class A&&\\
\hline
SDSSJ081257.252+181914.77&2.10&$3.31\pm0.146$\\
SDSSJ112343.182-010315.47&2.09&$1.57\pm0.154$\\
SDSSJ114753.301+3131 2.99&2.04&$1.57\pm0.278$\\
\hline
Class B&&\\
\hline
SDSSJ005018.623+050132.50&1.28&$0.81\pm0.130$\\
SDSSJ013556.390-001631.83&1.02&$0.93\pm0.104$\\
SDSSJ093323.128-012307.61&1.13&$1.13\pm0.142$\\
SDSSJ133755.789+402150.20&1.15&$0.97\pm0.135$\\
SDSSJ144437.728-013625.50&5.68&$5.33\pm0.146$\\
SDSSJ145924.058+035622.40&3.31&$3.07\pm0.145$\\
SDSSJ160747.246+162123.67&3.43&$3.17\pm0.138$\\
SDSSJ161404.729+042122.83&10.21&$9.89\pm0.154$\\
SDSSJ163414.493+231737.47&3.38&$3.40\pm0.148$\\
\hline
\end{tabular}
\end{center}
\label{tab:radio}
\end{table}

\subsection{Mid-Infrared: Wide-Field Infrared Survey Explorer (WISE)}
\label{sec:WISE}

WISE \citep{2010AJ....140.1868W} observed the entire sky
twice in four bands centered at 3.4$\mu$m, 4.6$\mu$m, 12$\mu$m and
22$\mu$m (W1, W2, W3 and W4) with 5-$\sigma$ sensitivity of 0.08,
0.11, 1 and 6 mJy and angular resolution 6.1$''$, 6.4$''$, 6.5$''$ and
12.0$''$ respectively.  At the typical redshift ($z = 2.5$) of our
sample, the four WISE bands correspond to  rest-frame 1$\mu$m, 1.3$\mu$m,
3.4$\mu$m, and 6.3$\mu$m, respectively.  Quasar SEDs show a break at
$\sim 1\mu$m, longwards of which hot dust can dominate the SED
\citep[e.g.,][]{2006ApJS..166..470R}, so only the relatively
low-sensitivity W3 and W4 bands can tell us about dust emission in our
objects.  

Obscured quasars are expected to have a high ratio of rest-frame 
IR-to-optical emission. The IR light is produced by thermal 
emission of the obscuring material, so it should be present 
in both obscured and unobscured quasars, whereas the 
optical light is strongly suppressed in the latter. Unfortunately, 
the high redshifts of our candidates mean that WISE probes 
the rest-frame near-IR emission, which is not produced by the 
obscured material in large amounts and is thus not particularly 
strong in the SED of unobscured quasars 
(Figure~\ref{fig:COSMOS_SED}). Furthermore, in obscured quasars the 
rest-frame near-IR may also be affected by extinction. 
The exception is the 22$\mu$m\ band 
of WISE, which probes close to the peak of the infrared emission, 
but in most cases the WISE catalog in this band is not sensitive
 enough for detecting our sources. 

We accessed the WISE Source Catalog using the NASA/IPAC Infrared
Science Archive (IRSA)\footnote{\tt
http://irsa.ipac.caltech.edu/cgi-bin/Gator/nph-dd} and conducted a
search within 2$''$ of the positions of our class A and class B quasar
candidates.  Forty of our class A objects (27.6\%) and 62 (20.2\%) of
our class B objects appeared in the WISE catalog.  While
\citet{2012AJ....144...49W} found that more than 85\% of optically
bright quasars ($i < 19.5$) in DR7 had a WISE match, less than 50\% of
those with $i > 20.5$ were matched, consistent with our result.  While
the S/N is typically low in bands W3 and W4 (due to the decreased
sensitivity of these bands), 27 class A sources and 29 class B sources
have a S/N above 5 in both bands W1 and W2 (only objects that were
detected with a S/N $> 5$ in at least one band are included in the
WISE All-Sky Release Catalog).  These sources are listed in
Tables~\ref{tab:classA_WISE} and \ref{tab:classB_WISE}.  WISE
magnitudes are given on the AB system \citep{1983ApJ...266..713O}.  Note that many errors are
listed as ``null'', indicating that the source is undetected in that
band; the values listed represent $2\,\sigma$ upper limits on the
flux.  

There are only nine Class A objects
with S/N above 5 in W3, and only two in W4.  Ross et al. (in
preparation) will discuss the properties of the most extreme W4
detections, with colors approaching those of the dust-obscured galaxy
population discovered by \citet{2008ApJ...677..943D}, among the SDSS
and BOSS quasars.  While none of these most extreme sources are among
our Type II quasar candidates, several of them show very weak
continuum and broad emission lines.  While their line widths are above
our nominal 2000 \kms\ limit, their high IR-to-optical ratios suggests that they are 
obscured quasars.  We will explore relaxing our rigid line-width
criterion in future work.  

\begin{figure}
\begin{center}
\includegraphics[width = 12cm]{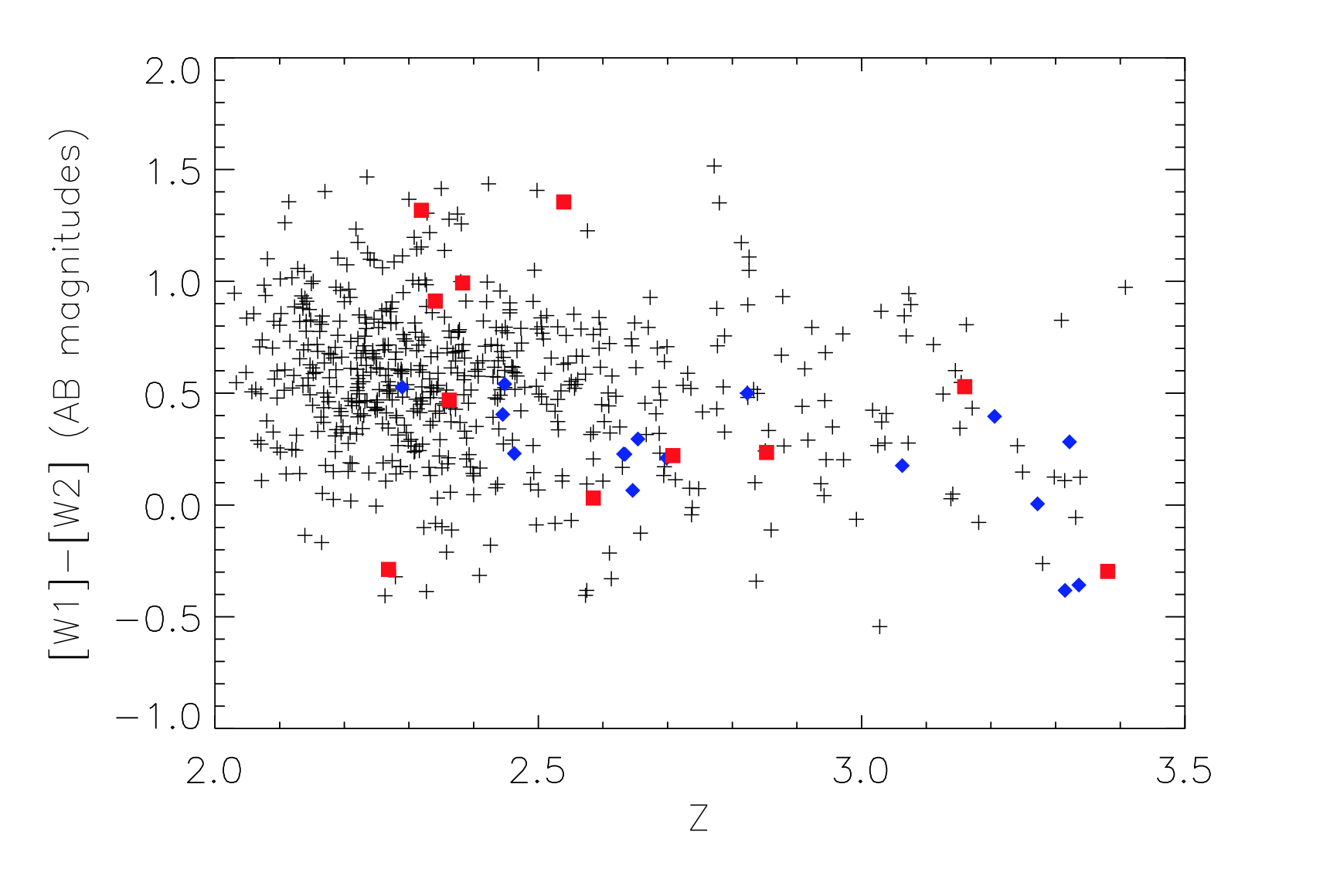} 
\end{center}
\caption{WISE color  (W1-W2; i.e., [3.4]-[4.6])  as  a function  of
  redshift  for  candidate Type  II  quasars below redshift 3.5;  red squares are  those
  objects detected above 5$\sigma$ in  both bands, while the blue
  diamonds are sources  detected below 5$\sigma$ in one  or both bands.
  The  gray crosses  are a sample of Type I quasars  from DR9  with
  a  similar distribution of $i$-band magnitude with WISE
  detections, and thus similar photometric errors.  The distribution
  of colors is indistinguishable between the two samples with a K-S test.}
\label{fig:WISE_colours} 
\end{figure}

Figure~\ref{fig:WISE_colours} shows the WISE $[3.6]-[4.5]$ color of
Class A sources as a function of redshift; a sample of DR9 BOSS
quasars with similar $i$-band magnitudes is included for comparison.  The
colors of both samples become bluer at redshift $z \sim 3.5$, when
H$\alpha$ enters the W1 band (although this redshift range is not shown
in Figure \ref{fig:WISE_colours}).  A K-S test says the color distributions
of the two are the same at the 99.99$\%$ level for those objects with
redshift less than 3.5; there is no evidence that the Type II objects
are redder in this color than unobscured quasars of similar
brightness.  

\begin{figure}
\begin{center}
\includegraphics[width=12cm]{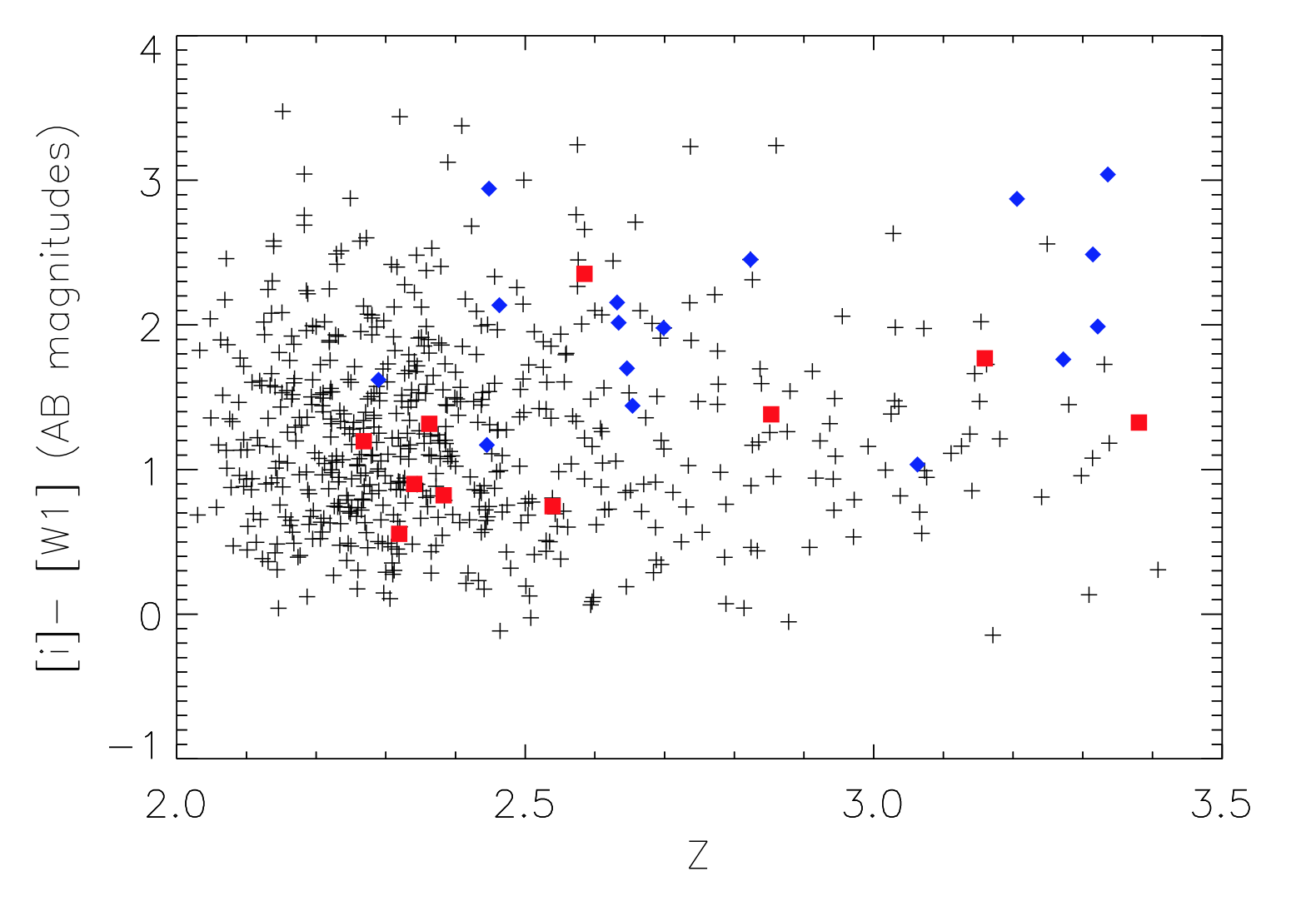}
\end{center}
\caption{The color of Type II candidates between the SDSS $i$ and W1 bands,
as a function of redshift; symbols are as in
Figure~\ref{fig:WISE_colours}.  The distributions of colors of the
Type II and Type I samples are indistinguishable with a K-S test.}
\label{fig:WISE_SDSScolor}\end{figure}

Similarly, Figure~\ref{fig:WISE_SDSScolor} measures a color between the
SDSS and WISE bands.  A K-S test shows that the color distributions
are the same at the $99.9\%$ confidence level. Much of the scatter in this diagram
and Figure~\ref{fig:WISE_colours} is due to measurement errors both in
SDSS and WISE; these detections are often near the flux limit of both
surveys.  

\begin{table}
\caption{All WISE matches in our Class A sample.  Values are given in AB magnitudes.}
\begin{center}
\begin{tabular}{|c|c|c|c|c|c|c|c|c|}
\hline
SDSS Name & W1 & W1err & W2 & W2err & W3 & W3err & W4 & W4err\\
\hline
\hline
SDSS0046+0005 & 19.77 & 0.19 & 18.92 & 0.22 & 16.57 & null & 14.44 & 0.25 \\
SDSS0146+1211 & 19.60 & 0.12 & 19.42 & 0.19 & 17.24 & null & 15.28 & null \\
SDSS0154+0157 & 20.40 & 0.21 & 20.44 & 0.47 & 18.13 & null & 15.63 & null \\
SDSS0206+0104 & 19.61 & 0.12 & 19.31 & 0.18 & 17.49 & 0.29 & 15.83 & 0.47 \\
SDSS0232+0028 & 19.49 & 0.10 & 18.96 & 0.12 & 17.95 & 0.44 & 16.07 & 0.53 \\
\hline
\end{tabular}
\end{center}
\label{tab:classA_WISE}

Only a portion of this table is shown here to demonstrate its form and
content. A machine-readable version of the
full table will be published online.  
\end{table}

\begin{table}
\caption{All WISE matches in our Class B sample.  Values are given in AB magnitudes.}
\begin{center}
\begin{tabular}{|c|c|c|c|c|c|c|c|c|}
\hline
SDSS Name & W1 & W1err & W2 & W2err & W3 & W3err & W4 & W4err\\
\hline
\hline
SDSS0010+0003 & 19.46 & 0.15 & 18.70 & 0.13 & 17.38 & 0.35 & 15.39 & 0.51 \\
SDSS0011$-$0008 & 19.29 & 0.12 & 19.67 & 0.34 & 17.51 & 0.45 & 15.10 & null \\
SDSS0201+0134 & 20.16 & 0.19 & 20.06 & 0.36 & 17.77 & null & 15.38 & null \\
SDSS0232$-$0812 & 18.46 & 0.05 & 18.38 & 0.07 & 17.17 & 0.20 & 16.00 & 0.50 \\
SDSS0234$-$0754 & 20.55 & 0.23 & 19.42 & 0.16 & 17.62 & 0.33 & 16.10 & null \\
\hline
\end{tabular}
\end{center}
\label{tab:classB_WISE}

Only a portion of this table is shown here to demonstrate its form and
content. A machine-readable version of the
full table will be published online.  
\end{table}%

\subsection{Mid-Infrared: Spitzer MIPS-24}
\label{sec:spitzer}

Only a handful of objects are detected in the long wavelength filters
of WISE.  In order to obtain an independent estimate of the bolometric
luminosities of our sources, we have searched the Spitzer Space
Telescope archive for serendipitous coverage of our Class A and Class
B samples. Using the Spitzer Heritage
Archive\footnote{\tt http://sha.ipac.caltech.edu/applications/Spitzer/SHA/},
we searched for Multiband Imaging Photometer (MIPS;
\citealt{riek04}) 24\micron\ data covering the positions of our Class
A and Class B sources. 
MIPS-24 is fairly close to the peak of the spectral
energy distribution for normal unobscured quasars, whereas 
serendipitous MIPS observations at longer wavelengths (70 and
160\micron) are too shallow to yield any detections of our sources. 

We found 13 sources that are covered by MIPS-24 observations.   
Several objects are covered by multiple exposures, which we coadded to increase the effective
exposure time at the positions of our targets. We then conducted aperture photometry
at the SDSS positions (accurate to 0.1\arcsec; \citealt{2003AJ....125.1559P}), using a circular aperture
with a radius of 12\arcsec\ and a background annulus between 12\arcsec\ and
18\arcsec. 

The photometric measurements are presented in
Table~\ref{table:spitzer}. Five out of 13 sources are detected at the
$4\,\sigma$ level or greater; another four sources are marginally
detected. We used the observed Spitzer fluxes to calculate the
rest-frame monochromatic luminosity at 6\micron. At the redshifts of
our detected sources, an observed wavelength of 24\micron\ corresponds
to rest-frame 5.9$-$7.3\micron; we k-correct to 6\micron\ using
$F_{\nu}\propto \nu^{-1.09}$ from the average Type I spectral energy
distribution by \citet{2006ApJS..166..470R}. The bolometric
luminosities of our objects can be estimated by multiplying the
monochromatic luminosities at 6\micron\ by a factor of 8, again taken
from \citet{2006ApJS..166..470R}.

Thus approximately half of the objects covered by the MIPS-24 data are detected at
the level 0.5$-$1 mJy, with detection luminosities in the range $\nu
L_{\nu}[6\micron]=10^{45.4-45.9}$ erg s$^{-1}$ and estimated bolometric luminosities
$10^{46.3-46.8}$ erg s$^{-1}$.  Thus these objects are indeed very
luminous quasars. 

\begin{table}
\caption{Spitzer 24\micron\ Data}
\label{table:spitzer}
\begin{center}
\begin{tabular}{lrlll}
Object & Flux Density (mJy) & Class & $\log(\nu L\nu/(\rm erg s^{-1}))$ (6\micron)\\
\hline
{\bf Detections:}\\
SDSSJ095819.35+013530.5  &$    0.76  \pm     0.10 $&A & 45.89 \\
SDSSJ010554.41+011326.9  &$    0.85  \pm     0.16 $&B & 45.94 \\
SDSSJ122353.62+050321.0  &$     1.01 \pm     0.12 $&B & 45.77 \\
SDSSJ135136.57+381642.8  &$    0.42  \pm     0.11 $&B & 45.39 \\
SDSSJ142610.76+341738.9  &$     1.07 \pm     0.12 $&B & 45.72 \\
\\
{\bf Marginal detections and non-detections:}\\
SDSSJ021834.53-033518.5  &$   0.21\pm     0.06$& A \\
SDSSJ161447.98+354221.2  &$  -0.11  \pm    0.14 $&A \\
SDSSJ020546.33+013907.6  &$  -0.10 \pm     0.14 $&B \\
SDSSJ022429.14$-$024807.7  &$   -0.28 \pm    0.08 $&B \\
SDSSJ115840.06+014335.3  &$    0.15 \pm    0.06 $&B \\
SDSSJ121326.70+062922.8  &$    0.24 \pm    0.07 $&B \\
SDSSJ124219.55+413720.6  &$    0.21\pm    0.08 $&B \\
SDSSJ141649.80+365012.2  &$    0.87 \pm     1.04 $&B \\
\hline
\end{tabular}
\end{center}
\end{table}

\subsection{X-Ray: XMM-Newton, Chandra and ROSAT}
\label{ssec:xray}

Luminous X-ray emission is nearly ubiquitous in unobscured quasars 
\citep[e.g.][]{2004ASSL..308...53M,2008ApJ...685..773G}, and high-quality X-ray spectra of these 
Type~II quasar candidates could usefully constrain their obscuration levels 
and/or intrinsic luminosities \citep[e.g.][]{2005ARA&A..43..827B,2010MNRAS.404...48V,
2011ApJ...738...44A,2011A&A...526L...9C,2012arXiv1205.0033J}. We therefore 
have searched the Chandra, ROSAT, and XMM-Newton archives for sensitive X-ray 
coverage of the sources in our sample. Sources having sensitive X-ray coverage 
were individually inspected in X-ray and SDSS images to assess critically the 
reliability of putative X-ray detections and X-ray/SDSS
associations. This screening identified five X-ray detected objects
and another that is likely X-ray detected (SDSSJ160554.89+202729.87); 
the X-ray 
properties of these sources are listed in Table~\ref{tab:xray}.
All of these sources were serendipitous detections rather than the targets
of their respective observations. In addition, five sources 
(SDSSJ023337.89+002303.69, SDSSJ023359.27+005925.83, SDSSSJ120355.19+014348.98, 
SDSSJ144227.31$-$004725.04, and SDSSJ144441.05$-$001343.44) 
are undetected in X-rays in observations of generally comparable
sensitivity to those that yielded  
the detections in Table~\ref{tab:xray}. 

\begin{table}
\caption{X-Ray detections of Type II quasar candidates by Chandra, ROSAT, and XMM-Newton}
\begin{tabular}{|l|l|l|r|r|r|l|}
\hline
SDSS Coordinate&Facility&Observation&Screened Effective&Energy&Observed Flux&Notes\\
&&ID&Exposure (ks)&Band (keV)&($10^{-15}$ erg s$^{-1}$ cm$^{-2}$)&\\
\hline
\hline
014841.83+055024.19&XMM-Newton&0112551501&10.6&0.2-12&16&2XMM\\
082726.58+262803.00&XMM-Newton&0103260601&11.1&0.2-12&23&2XMM\\
082726.76+214557.08&Chandra&10268&10.0&0.3-8&31&\\
095819.35+013530.52&XMM-Newton&0302353301&6.3&0.2-12&24&2XMM \\
142610.77+341738.85&Chandra&7388&4.5&0.5-8&4.8&$^1$ \\
160554.89+202729.87&ROSAT&rp600588n00&16.1&0.5-2&8.6&$^2$ \\
\hline
\end{tabular}
\label{tab:xray}

{Three of the sources are included in the second XMM-Newton
serendipitous source catalogue (2XMM;
\citealt{2009A&A...493..339W}). The screened effective exposure time
corrects for the effects of vignetting and for time lost to flares.\\
$^1$Two-photon detection; in NDWFS Bootes field,\\
$^2$ROSAT identification likely correct but needs verification.
}
\end{table}%

The sources with X-ray detections in Table~\ref{tab:xray} have 2--99
counts; five of the six sources have 45 or fewer counts. Most also
have high backgrounds in the source detection cell owing to large
off-axis angles. Thus, unfortunately, none of these sources has
sufficient S/N for useful X-ray spectroscopy that could constrain, via
X-ray spectral shape, absorption and intrinsic luminosity. Additional
X-ray observations are required to perform such
spectroscopy.  Alternatively, once independent multiwavelength luminosity
indicators are available for these sources (e.g., \oiii\ 5008~\AA\ and
far-infrared luminosities), these can be correlated with even basic
X-ray emission measurements to assess obscuration levels
\citep[e.g.][]{2010MNRAS.404...48V,2012arXiv1205.0033J}.

\subsection{Cosmic Evolution Survey}
\label{ssec:cosmos}

The Cosmic Evolution Survey
(COSMOS)\footnote{\tt http://cosmos.astro.caltech.edu/index.html} covered
2 deg$^2$ with the Hubble Space Telescope (HST) Advanced Camera for
Surveys \citep[ACS;][]{2007ApJS..172..196K}, the VLA \citep{2004AJ....128.1974S}, 
the Spitzer Space Telescope \citep{2007ApJS..172...86S}, XMM-Newton \citep{2007ApJS..172...29H} and
other facilities. One of our sources, 
SDSSJ095819.35+013530.5 at
$z=3.0554$ (hereafter SDSS0958+0135), lies in the COSMOS field. This object therefore has
extensive multi-wavelength coverage and an exquisitely measured
SED. The optical and NIR photometry from 15
bands is available via the Infrared Science Archive (IRSA), and so is the Spitzer IRAC 
(3.6$\mu$m-8$\mu$m) and MIPS-24$\mu$m\ photometry (as we saw in
\S~\ref{sec:spitzer}).
The source is not in the MIPS-70$\mu$m\ catalog of
COSMOS. We co-added all 14 available MIPS-70$\mu$m\ exposures, conducted
aperture photometry and obtained a tentative detection with a flux of
$1.13\pm 0.7$ mJy.

\begin{figure}
\begin{center}
\includegraphics[width = 12cm,angle=270]{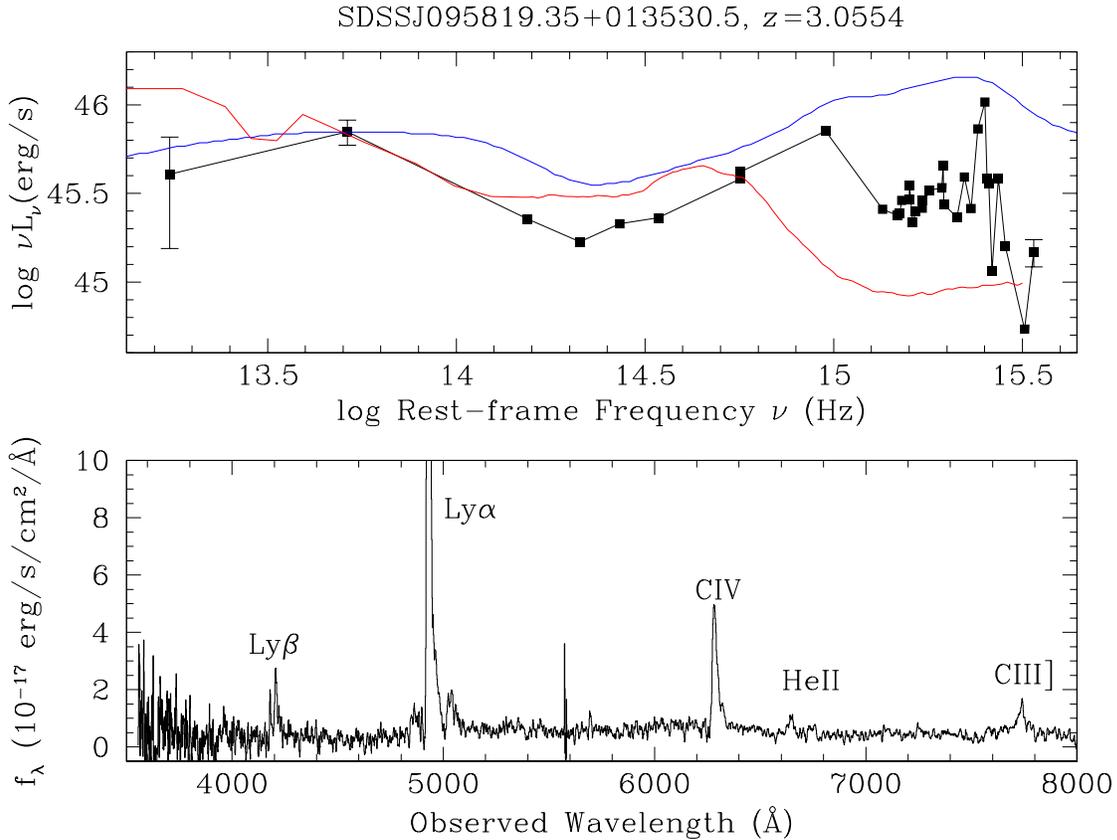}
\end{center}
\caption{{\it Upper:} The spectral energy distribution (SED) of
SDSS0958+0135, which lies in the COSMOS field, including data from the
Spitzer Space Telescope and a variety of ground-based facilities.  The
peak at $\log \nu = 15.4$ is due to Ly$\alpha$ emission.  For
comparison, the blue line is the composite
  Type I SED from \citet{2006ApJS..166..470R} and the red line is the 
  mean low-redshift Type II quasar SED \citep{Zakamska08}.  All
  curves are normalized to a rest-frame frequency of $10^{13.7}$ Hz (6
  $\mu$m, corresponding to the 24 $\mu$m detection with
  Spitzer).   {\it Lower:} The BOSS spectrum of this source with expanded
  vertical and horizontal axes (the peak of the Ly$\alpha$ line is well off-scale),
  with emission lines marked. The \civ\ line is markedly asymmetric,
  perhaps indicating self-absorption on the blue side.}
\label{fig:COSMOS_SED}
\end{figure}

In Figure \ref{fig:COSMOS_SED} we use multi-band data from the COSMOS
survey to construct the SED of this object and compare it with those
of a Type I quasar template from \citet{2006ApJS..166..470R} and a
Type II (obscured) quasar template derived from Spitzer data
\citep{Zakamska08} of low-redshift luminous optically-selected
obscured quasars \citep{2003AJ....126.2125Z}. The templates have been
arbitrarily normalized to match the observed SED at rest-frame
6$\mu$m\ (as measured by MIPS-24$\mu$m).

The SED of SDSS0958+0135 shows several noteworthy features. The
ultraviolet to infrared flux ratio is significantly lower than in
unobscured quasars (and the high equivalent width of Ly$\alpha$ in
emission contributes significantly to the ultraviolet flux). However, the
ultraviolet emission in this source is not as suppressed as it is in
low-redshift obscured quasars. This feature suggests that SDSS0958+0135
is reddened by moderate amounts of dust, perhaps in its host
galaxy. Extinction with $A_V=0.3-0.6$ mag would
suppress the ultraviolet continuum by a factor of 3, but would leave longer
wavelengths much less affected. Another possibility is that the object
is strongly absorbed, and the observed continuum in the ultraviolet
is produced by quasar light that is scattered off the material in the
host galaxy and reaches the observer, even though the central engine
itself is not visible along our line of sight.  However, matching the
observed optical SED would require $\sim$ 10\% 
scattering efficiency, which is quite extreme. 
Future polarization measurements of the kind discussed in the 
next section will help distinguish these two possibilities. 

The SED of SDSS0958+0135 appears to peak at relatively
short wavelengths -- around rest-frame 6$\mu$m\ -- declining to
longer wavelength, which is similar to the shape of the SED of
unobscured quasars, and unlike low-redshift obscured quasars
whose SEDs tend to show more cold dust emission. The shape of the
mid-infrared SED is consistent with that of a reddened quasar, since
modest amounts of reddening applied to a Type I spectrum would not
affect mid-infrared wavelengths. Another possibility is a strongly
obscured source, but with a compact and hot obscuring region rather
than an extended colder one. The observed MIPS-24 flux of 0.76 mJy
corresponds to a monochromatic luminosity of $\nu
L_{\nu}[6\mu\rm m]=7.7\times 10^{45}$ erg sec$^{-1}$, as we saw in
\S~\ref{sec:spitzer}. The similarity of the
mid-infrared SED of SDSS~J0958+0135 to that of Type I quasars enables
us to use bolometric corrections derived by
\citet[][a factor of 8 at this wavelength]{2006ApJS..166..470R},
leading to $L_{\rm bol}=6\times 10^{46}$ erg sec$^{-1}$ in this
source.  This is clearly a very luminous quasar.

As part of the COSMOS survey, this source was observed in the 0.5-2
keV band by XMM-Newton; the inferred luminosity in the 2-8 keV
rest-frame is $9.2 \times 10^{43}$ erg sec$^{-1}$ (not shown in
Figure~\ref{fig:COSMOS_SED}).  This object was also observed by the
Hubble Space Telescope in the ACS/F814W filter, just redward of the
Ly$\alpha$ emission, for a full orbit (2028 seconds on-source). The
object is unresolved in the HST image; neither the emission from the
host galaxy nor extended scattered light are detected.  At this
redshift, $0.1''$, a reasonable upper limit on the size of the object
in the ACS data, corresponds to a physical extent of roughly 1 kpc,
suggesting a very compact object indeed.  This is not surprising if
the light is dominated by the direct (albeit reddened) light from the
quasar itself. If, on the other hand, the object is a highly obscured
quasar, then the limit on the size of the rest-frame ultraviolet
emission suggests that the scattering interstellar medium in the host
galaxy is compact. This model would be consistent with recent
findings that massive galaxies at high redshifts tend to have small
effective sizes \citep{wuyt10}.

\subsection{Optical polarization}
\label{sec:polarization}

Optical polarimetry is a classical test of the obscuration-based
unification model of active galactic nuclei \citep{antonucci85}. Even if
the direct line of sight to the nucleus is blocked by obscuration,
quasar light can escape along an unobscured direction (or sometimes
more than one, \citealt{schm07}), scatter off free electrons or dust
particles in the interstellar medium of the host galaxy, and reach the
observer. Since scattered light is polarized, optical polarimetry and
spectropolarimetry are uniquely sensitive to the scattered component.

This process occurs both in obscured and unobscured quasars, but in
the latter the scattered component is diluted by the emission from the
quasar itself, so the typical levels of polarization in unobscured
quasars is 0.5\% \citep{berr90}. In obscured or heavily reddened
quasars with large enough column density along the line of sight, the
scattered component dominates over the direct emission, and the
typical levels of polarization are much higher -- a few per cent
\citep{tran95, 2002ApJ...569...23S, smit03, 2005AJ....129.1212Z}, reaching $\ga 20$\% in several
exceptional cases \citep{1995ApJ...450L...1H,smit00,
2005AJ....129.1212Z}. Thus high levels of
polarization are strongly suggestive of an obscured active nucleus.  The
spectrum of the polarized component can be that of an unobscured
quasar, showing a blue continuum and broad permitted emission lines
\citep{antonucci85, 2005AJ....129.1212Z}. 

\begin{figure}\centering\includegraphics[width=16cm]{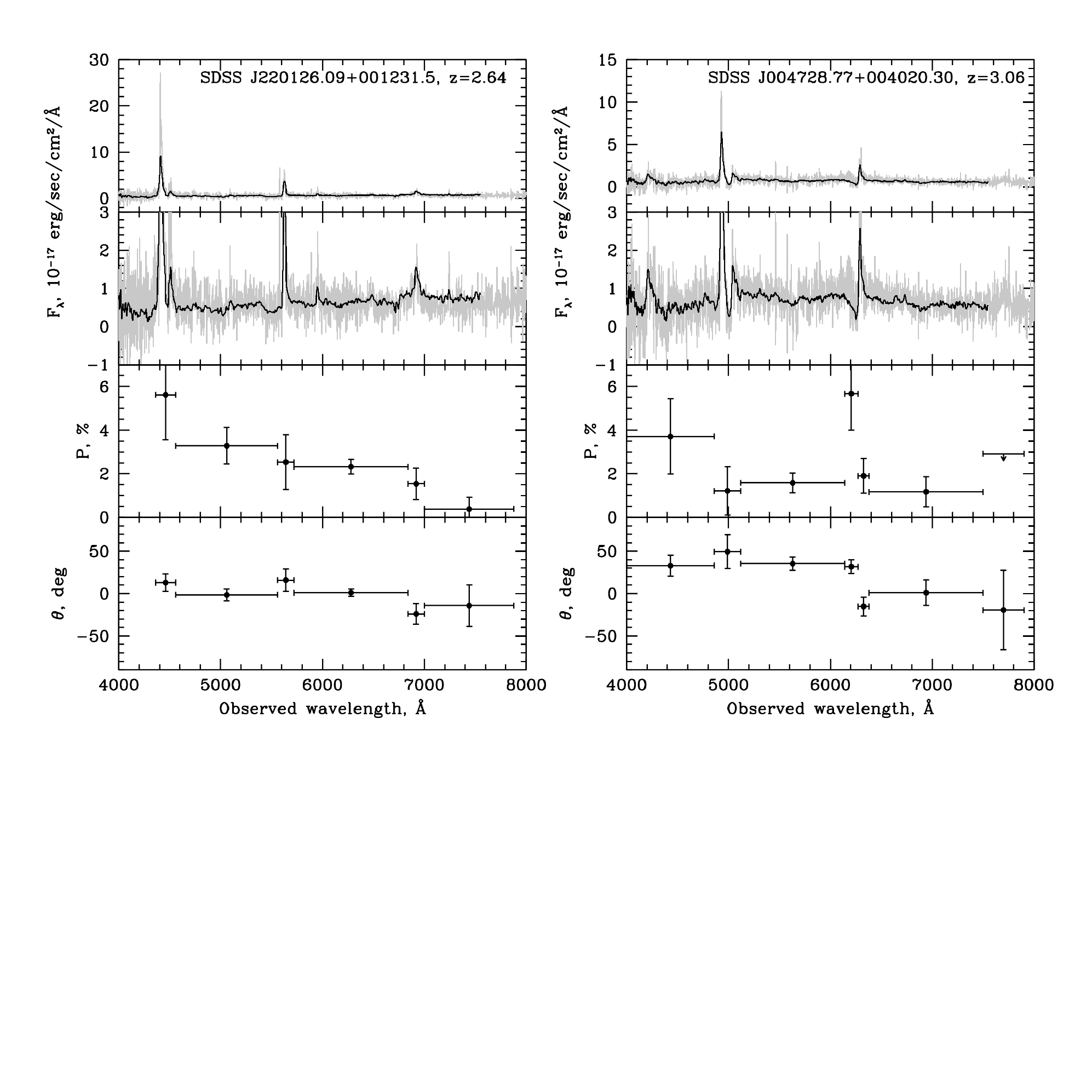}
\caption{
Results of spectropolarimetric observations of two candidate obscured
quasars with SPOL on MMT. In the top panels, we show the BOSS spectra
in gray and MMT spectra in black (both the overall spectra and an
expanded view to highlight weak features are shown). The MMT spectra were taken
in non-photometric conditions and have been scaled up by a factor $\sim 1.5$ to
line up with the SDSS continuum. In the bottom panels we show the
fractional polarization measured using SPOL observations, in percent,
and the polarization position angle in degrees E
of N. These have been binned over interesting wavelength regions
(continuum and lines) to build up S/N.}
\label{pic:mmt}
\end{figure}

In 2012 September we observed two high-redshift Type II quasar
candidates from our Class A sample with the CCD Spectropolarimeter (SPOL;
\citealt{schm92}) on the 6.5m MMT using the National Optical
Astronomical Observatory time allocation in sub-optimal weather
conditions. SDSSJ220126.11+001231.5 was observed over 4
hours, and SDSSJ004728.77+004020.3 over 2 hours, with a
1.1\arcsec\ slit and low-resolution grating, resulting in wavelength
coverage of 4000-8000~\AA\ and spectral resolution 19~\AA. 
Because the sources are faint ($i=20.7$ and $i=21.2$ mag,
respectively), we bin the data in wavelength,
allocating bins separately for continuum-dominated and
emission-line-dominated regions. The combined results of the MMT
observations are shown in Figure~\ref{pic:mmt}.  Unfortunately, this
binning and the limited S/N mean that we are not able to measure the
width of the lines in polarized light.  

When averaged over the high S/N continuum region of 5000$-$7000~\AA,
SDSS2201+0012 is polarized at a level of $1.9\pm 0.3$\%, with position
angle $176^\circ \pm 4^\circ$ (East of North). This polarization is
significantly higher than the typical polarization of unobscured
quasars (0.5\%), any instrumental systematics ($\la 0.1$\%), or
polarization that can be acquired due to the propagation of light
through the dust in the Galaxy ($<0.8$\%, \citealt{berr90}). There is
no evidence that the narrow emission lines are polarized any less than
the surrounding continuum, indicating that the scattering medium is
distributed on scales that are larger than the narrow line region. For
the COSMOS source (\S~\ref{ssec:cosmos}), we concluded that the
scattering region had to be compact; additional spectrophotometry and
high-resolution imaging of the same sources
\citep{2006AJ....132.1496Z} will be valuable to determine
whether there is a contradiction here. 

The polarization fraction rises toward the blue part of the spectrum,
where it reaches $>$5\%. This change is either indicative of a
wavelength-dependent scattering mechanism or of a red unpolarized
component that dilutes blue polarized light \citep{2005AJ....129.1212Z}. Since
electrons scatter optical and UV light in a wavelength-independent manner, the first
possibility calls for dust particles to be primarily responsible for
scattering. This is not an uncommon occurrence, especially in
high-luminosity obscured or reddened quasars where scattering occurs
over a large fraction of the entire host galaxy
\citep{hine01, 2005AJ....129.1212Z, schm07} rather than being confined to
the highly ionized circumnuclear material
\citep{antonucci85}. However, the contribution of the unpolarized but
reddened quasar continuum can mimic the wavelength dependence of
the polarized fraction, and we cannot rule out either of these
explanations with the current data.  Rest-frame optical
(observed-frame near-infrared) observations will help distinguish
these scenarios by allowing us to determine the shape of the continuum
and thus the level of quasar obscuration. The high S/N MMT spectrum
reveals a hint of a broad shallow absorption feature blueward of \civ,
which is thought to come from the central region in AGN, suggesting
that the light from the central engine is not completely obscured.  In
any case, the high level of polarization seen in SDSS2201+0012
strongly suggests that it is either an obscured or a heavily reddened
quasar.

SDSS0047+0040 is not as highly polarized as SDSS2201+0012, with a
average polarization over 5000 to 7000\AA\ of $0.91\pm 0.35$\%
(position angle $28^\circ \pm 10^\circ$). The most striking feature in
this source is that the mini-absorption trough blueward of the \civ\
emission line is polarized at a level of $\sim 6$\%, or $>3\sigma$
higher than the continuum. This feature is similar to that seen in
broad absorption line quasars (e.g., \citealt{dipo11}). Indeed, in
these sources the absorption trough blocks the direct light from the
quasar and thus suppresses the quasar continuum contribution relative
to the scattered light emission produced on larger spatial scales,
increasing the level of polarization seen within the trough relative
to that seen at other wavelengths. We conclude that the modest level
of polarization of SDSS0047+0040 and the polarization increase within
the mini-trough suggest that some of the continuum in this source is
direct quasar light.  The definitive test whether these two objects
are obscured would be broad emission lines in the polarized spectrum.
This will require substantially higher S/N spectropolarimetry than we
have obtained.

\section{Discussion}
\label{sec:discussion}

We consider our objects to be candidate Type II quasars by drawing an
analogy with the defining characteristics of Seyfert II galaxies at
lower redshift: narrow permitted lines seen in the rest-frame
ultraviolet, and a relatively weak continuum, as quantified by high
equivalent widths of these lines.  The key questions, which we have 
yet to answer, are: (a) what are the intrinsic luminosities of these 
objects; and (b) what is the range of the obscuration we probe in this sample.  We
do know that at least some of these sources are highly luminous
($>10^{46}$ erg sec$^{-1}$), given
detections in the Spitzer MIPS 24\micron\ band (\S~\ref{sec:spitzer}). Although WISE 
data do not go out to long enough wavelengths to probe the peak of 
the IR SED, they indicate similarly high luminosities for a few 
other sources in this sample.  X-rays
are another probe of unobscured luminosity; observations in the 2-10
keV energy range sample rest-frame energies from 7 to 35 keV, where
even moderately Compton-thick sources should be detected.  We have serendipitous
detections of a handful of sources (\S~\ref{ssec:xray}), but targeted
observations with Chandra or XMM-Newton will yield new insights into
their true luminosities.  

Our objects have continuum absolute magnitudes of $-23$ and brighter
in the ultraviolet (Figure~\ref{fig:luminosity_dist}); this is too
luminous to be explained by the stellar continuum of the host galaxy
of an obscured quasar.   Indeed, the brightest unlensed galaxies known
at $z > 2$ are roughly $r=24$ \citep{Shapley11}, two magnitudes fainter than the
faintest objects in our sample.  Thus the continuum light that we are
seeing must be dominated by the nucleus itself.  There are three
possible explanations: (a) these sources are only modestly extincted ($A_V
\sim 0.5$, corresponding to 1.5 magnitudes of extinction at 1500~\AA\ for
an $R=3.1$ extinction law), (b) substantial amounts of quasar
continuum light are scattered by distributed dust or electrons, as is
seen in lower-redshift Type II quasars, or (c) the line of sight to
the central region is heavily obscured, but the covering fraction is
not complete, allowing some amount of unobscured continuum through.
The relatively blue continuum and the broad bases seen on the emission
lines in the composite spectrum (Figure~\ref{fig:coadd}) are
consistent with all three of these hypotheses, and the rough similarities
in the broad-band colors and SEDs of these objects and Type I objects
are supportive of the scattering and partial covering hypotheses.  We
have polarization data on only two objects to date
(\S~\ref{sec:polarization}); the fact that both objects are strongly
polarized suggests extincted objects with substantial scattering.  It
may well be that our sample is heterogeneous, with some more extincted
than others.  With this in mind, it is dangerous to draw conclusions
about the whole population from just a few objects.  

Lower-redshift obscured quasars have been identified as such by their
narrow Balmer emission lines.  Near-infrared spectroscopy allows
measurement of these lines directly in our high-redshift objects, to
determine whether the permitted lines remain narrow at wavelengths
where extinction due to dust is significantly smaller.  We will report
observations in the near-infrared in a paper in preparation;
consistent with the modest extinction hypothesis, many of our sources
do show evidence for broad Balmer lines. 
The near-infrared data will also allow us to measure the luminosity in
the \oiii5008\AA\ emission line, which has been used as a proxy for
bolometric luminosity. 

  If the continua in these objects include a contribution from
scattering, we might detect extended scattered emission in
high-resolution images, as is seen in lower-redshift Type II quasars
\citep{2006AJ....132.1496Z}.  The COSMOS object is unresolved in HST
images, but we have an ongoing HST program to image six sources from
our survey; the results will be presented in a future paper. 

  If the continuum we are seeing from these objects is dominated by
scattered light on kpc scales, one would not expect to see variability
on human timescales.  A number of our objects fall on the Equatorial
Fall Stripe in the Southern Galactic Cap (``Stripe 82''), which was
repeatedly imaged during the SDSS \citep{DR7}, allowing a search for
such variability.  

  In future work, we hope to use this sample to quantify the relative
numbers of obscured and unobscured quasars as a function of redshift
and luminosity.  This is a challenging task, given the uncertainties
we have just discussed about the amount of extinction and the true
luminosities of our sources.  However, these objects are not extremely
rare: the 145 sources in our Class A sample are drawn from SDSS DR9,
which represents only the first year of BOSS data taking, or about 1/3
of the final BOSS survey.  We can expect the sample to triple in size
by the time of the final BOSS data release in late 2014.  Future
wide-angle spectroscopic surveys on larger telescopes, such as those
to be carried out by the Prime Focus Spectrograph being built for the
Subaru Telescope \citep{2012arXiv1206.0737E}, may reveal substantially
more such objects.

\section{Conclusions}
\label{sec:conclusions}

We have identified a sample of candidate Type II quasars at redshifts
between 2.0 and 4.3 from the spectra of the Baryon Oscillation Spectroscopic
Survey.  They are characterized by strong narrow (FWHM $< 2000$ \kms)
\lya\ and \civ\ emission lines of high equivalent width.  Our sample
includes 145 ``Class A'' objects, plus an additional 307 ``Class B''
objects whose classification is less certain.  Our main conclusions are as
follows: 
\begin{itemize} 
\item These objects have continuum absolute magnitudes of $-23$
and brighter, suggesting that the quasar continuum is only modestly
extincted, that the extinction is patchy, or that the quasar continuum is strongly scattered.
\item A composite high S/N spectrum shows broad bases (FWHM$\sim3500$
\kms) in many emission lines, suggesting modest extinction or
substantial scattering of light from the central engine. 
\item The distribution of broad-band colors of these objects from the
rest-frame ultraviolet to 1\micron\ are
consistent with those of unobscured quasars at the same redshift. 
\item The ratios of the strengths of the \civ, \heii, and \ciii\ emission
 lines are distinctly different from those of other classes of Type II
 AGN at high redshift, suggesting a higher ionization parameter
 and a lower metallicity than these other samples.
\item Many of the objects show significant self-absorption in the \lya\ and \civ\ emission 
lines, often blue-shifted relative to the rest-frame peaks of the lines, 
suggesting substantial absorbing gas and outflows.
\item Serendipitous observations of a dozen objects at 24\micron\ with
the Spitzer Space Telescope imply bolometric luminosities above
$10^{46}$ erg s$^{-1}$.  
\item Polarization measurements of two objects suggest that there is
a significant scattered component to the continuum.  
\item Further insights into the intrinsic luminosities, obscuration,
and physical nature of these sources will require additional
X-ray, near- and mid-infrared, and spectropolarimetric observations. 
\end{itemize}

\section*{Acknowledgments}

We thank Kevin Hainline and Dan Stern for useful discussions and the
use of their spectra, and Gordon Richards and Joe Hennawi for useful
comments on an early draft of the paper.  

Funding for SDSS-III has been provided by the Alfred P. Sloan
Foundation, the Participating Institutions, the National Science
Foundation, and the U.S. Department of Energy Office of Science. The
SDSS-III web site is {\tt http://www.sdss3.org/}. 

SDSS-III is managed by the Astrophysical Research Consortium for the
Participating Institutions of the SDSS-III Collaboration including the
University of Arizona, the Brazilian Participation Group, Brookhaven
National Laboratory, University of Cambridge, Carnegie Mellon
University, University of Florida, the French Participation Group, the
German Participation Group, Harvard University, the Instituto de
Astrofisica de Canarias, the Michigan State/Notre Dame/JINA
Participation Group, Johns Hopkins University, Lawrence Berkeley
National Laboratory, Max Planck Institute for Astrophysics, Max Planck
Institute for Extraterrestrial Physics, New Mexico State University,
New York University, Ohio State University, Pennsylvania State
University, University of Portsmouth, Princeton University, the
Spanish Participation Group, University of Tokyo, University of Utah,
Vanderbilt University, University of Virginia, University of
Washington, and Yale University.  RA and MAS acknowledge the support
of NSF grant AST-0707266, and JEG and NZ acknowledge the support of Alfred
P. Sloan Foundation Fellowships.  NZ is also supported by the Theodore Dunham,
Jr., Grant of the Fund for Astrophysical Research. WNB acknowledges
the support of NASA ADAP grant NNX10AC99G and NSF
  grant AST-1108604.

This research has made use of the NASA/IPAC Infrared Science Archive,
which is operated by the Jet Propulsion Laboratory, California
Institute of Technology, under contract with the National Aeronautics
and Space Administration. 

\bibliography{type2}

\label{lastpage}

\end{document}